\documentclass[useAMS,usenatbib]{mn2e}
\usepackage{graphicx}
\usepackage{subfigure}

\newcommand{\gtrsim}{\let\ga}
\newcommand{\kms}{\,km\,s\ensuremath{^{-1}}}
\newcommand{\hmol}{H\ensuremath{_2}}
\newcommand{\fsiii}{\mbox{[S\,{\sc iii}}]}
\newcommand{\fsiv}{\mbox{[S\,{\sc iv}}]}
\newcommand{\fsix}{\mbox{[S\,{\sc ix}}]}
\newcommand{\foiii}{\mbox{[O\,{\sc iii}}]}
\newcommand{\ffeii}{\mbox{[Fe\,{\sc ii}}]}
\newcommand{\fcaviii}{\mbox{[Ca\,{\sc viii}}]}
\newcommand{\fpii}{\mbox{[P\,{\sc ii}}]}
\newcommand{\hi}{\mbox{H\,{\sc i}}}
\newcommand{\hei}{\mbox{He\,{\sc i}}}
\newcommand{\pab}{Pa\ensuremath{\beta}}
\newcommand{\brg}{Br\ensuremath{\gamma}}
\defcitealias{sb09}{Paper I}

\title[Feeding and Feedback in NGC\,4151. II]{Feeding versus Feedback in NGC\,4151 probed with Gemini NIFS. II. Kinematics}
\author[Storchi-Bergmann et al.]{T. Storchi-Bergmann$^1$$\thanks{E-mail: thaisa@ufrgs.br}$, R. D. Sim\~oes-Lopes$^{1, 4}$,   P. J. McGregor$^2$,
\newauthor  Rogemar A. Riffel$^1$,  T. Beck$^3$, and P. Martini$^4$\\
$^1$Instituto de F\'isica, Universidade Federal do Rio Grande do Sul, Av. Bento Gon\c calves 9500, 91501-970, Porto Alegre, RS, Brazil\\
$^2$Research School of Astronomy and Astrophysics, Australian National University, Cotter Road, Weston Creek, ACT\,2611, Australia\\
$^3$Gemini Observatory and Space Telescope Science Institute, 3700 San Martin Dr., Baltimore, MD 21218\\
$^4$Department of Astronomy and Center for Cosmology and Astroparticle Physics, The Ohio State University, Columbus, OH 43210}

\begin{document}

\date{Updated by Thaisa on October 9,  2009}

\pagerange{\pageref{firstpage}--\pageref{lastpage}} \pubyear{2009}
\maketitle
\label{firstpage}

\begin{abstract}

We have used the Gemini Near-infrared Integral Field Spectrograph (NIFS) to map the gas kinematics of the  inner $\sim$\, 200$\times$500\,pc of the Seyfert galaxy NGC\,4151 in the Z, J, H and K bands at a resolving power $\ge$\,5000 and spatial resolution of $\sim$\,8\,pc. The ionised gas emission is most extended along the known ionisation bi-cone at position angle PA=60--240$\degr$, but is observed also along its equatorial plane. This indicates that the AGN ionizes gas beyond the borders of the bi-cone, within a sphere with $\approx$\,1\arcsec\ radius around the nucleus. The ionised gas has three kinematic components: (1) one observed at the systemic velocity and interpreted as originating in the galaxy disk; (2) one outflowing along the bi-cone, with line-of-sight velocities between $-600$ and 600\,\kms\ and strongest emission at $\pm\,(100-300)$\,\kms;  (3) and another component due to the interaction of the radio jet with ambient gas.  The radio jet (at PA=75--255$\degr$) is not aligned with the NLR, and produces flux enhancements  mostly observed at the systemic velocity, suggesting that the jet is launched close to the plane of the  galaxy ($\sim$ plane of the sky).  The mass outflow rate, estimated to be $\approx$\,1\,M$_\odot$\,yr$^{-1}$ along each cone, exceeds the inferred black hole accretion rate by a factor of $\sim$\,100. This can be understood if the Narrow-Line-Region (NLR) is formed mostly by entrained gas from the circumnuclear interstellar medium by an outflow probably originating in the accretion disk. This flow represents feedback from the AGN, estimated to release a kinetic power of $\dot{E}\approx\,2.4\,\times10^{41}$\,erg\,s$^{-1}$, which is only $\sim$\,0.3\% of the bolometric luminosity of the AGN.

There is no evidence in our data for the gradual acceleration followed by gradual deceleration proposed by previous modelling of the [OIII] emitting gas. Our data allow the possibility that the NLR clouds are accelerated close to the nucleus (within 0$\farcs$1 -- $\approx$\,6\,pc) after which the flow moves at essentially constant velocity ($\approx\,600$\kms), being consistent with NIR emission arising predominantly from the interaction of the outflow with gas in the galactic disk.

The molecular gas exhiibits distinct kinematics relative to the ionised gas. Its emission arises in extended regions approximately perpendicular to the axis of the bi-cone and along the axis of the galaxy's stellar bar, avoiding the innermost ionised regions. It does not show an outflowing component, being observed only at velocities very close to systemic, and is thus consistent with an origin in the galaxy plane. This hot molecular gas may only  be the tracer of a larger reservoir of colder gas which represents the AGN feeding. 

\end{abstract}


\section{Introduction}

This is a continuing study of the narrow-line region (hereafter NLR) of the Seyfert galaxy NGC\,4151 using data obtained with the Gemini Near-infrared Integral Field Spectrograph (NIFS).
The data comprise spectra of the inner$~\approx\,200\,\times\,500\,$pc$^2$, at a spatial resolution of$~\approx\,8\,$pc at the galaxy, covering the wavelength range$~0.95$--$2.51\,\mu$m at a spectral resolving power R$\,\geq 5000$. In a previous paper \citep[][hereafter Paper\,I]{sb09} we have used these data to map the NLR intensity distributions of $14$ emission lines, as well as their ratios. The main results were the distinct flux distributions and physical properties observed for the ionised, molecular and coronal gas. The ionised gas is co-spatial with the ionisation bi-cone observed in the optical \foiii\ emission at position angle (hereafter PA)$~60\degr$ \citep{evans93,hutchings98}, and seems to trace the outflow along the bi-cone \citep{das05,crenshaw00a,hutchings99}. In the inner region, the NIR ionised-gas emission extends beyond the borders of the cone, which does not have a sharp apex as noted also by  \citet{kra08}. The \hmol\ molecular gas intensity distribution, on the other hand, avoids the region of the bi-cone and seems to originate in the galaxy disc, while the coronal gas emission is barely resolved. In another recent work \citep{riffel09b}, we have studied the unresolved nuclear continuum showing that its origin is emission by hot dust within $\approx$\,4\,pc from the nucleus -- as expected from the dusty torus postulated by the unified model \citep{antmi85}. The origin of this structure is probably a dusty wind that originates in the outer parts of the accretion disk. This wind is probably clumpy, as suggested by recent models \citep{elitzur06}, and necessary in order to allow the escape of  radiation along the equatorial plane of the bi-cone in order to ionize the gas and produce the observed flux maps of Paper\,I.

In the present paper we use the NIFS data to map the NLR kinematics of NGC\,4151. Although many papers have been devoted to such a study, most of them are based on long-slit spectroscopy obtained in the optical with the Hubble Space Telescope \citep[e.g.][]{winge97,hutchings98,crenshaw00a,das05}. The present study was performed in the near-IR, a waveband which is less affected by dust, and using integral field spectroscopy, which allows a full two-dimensional (hereafter 2D) coverage of the NLR kinematics. With our data and analysis we aim to examine previous claims of acceleration of the gas along the NLR, quantify the mass outflow rate and the corresponding feedback power, as well as investigate the origin of the NLR gas. As NGC\,4151 harbors the closest bright AGN, its NLR  is one of the best suited for this type of study.

Our approach for the analysis of the NLR kinematics of NGC\,4151 is as follows. First we map the centroid velocity and velocity dispersion of the ionised, molecular and coronal gas, obtained from fits to the emission-line profiles and compare our results with those of previous studies. Then we use an alternative approach to map the NLR kinematics, made possible by integral field spectrographs: a ``velocity tomography'' of the emitting gas, obtained by slicing the line profiles in velocity bins of$~60$\kms, producing channel maps which provide a clearer view of the velocity distribution in the different gas phases. We have grouped the channel maps together in a sequence of velocity bins, generating movies which can be recovered at the authors' website. \footnote{http://www.if.ufrgs.br/$\sim$thaisa/ifu\_movies/ngc4151}


We adopt in this paper the same distance to NGC\,4151 as in Paper\,I:$~13.3\,$Mpc, corresponding to a scale at the galaxy of$~65\,$pc\,arcsec$^{-1}$ \citep{mundell03}. The paper is organised as follows. In \S\,\ref{data} we provide a quick description of the data, in \S\,\ref{results} we present the centroid velocity and velocity dispersion measurements, in \S\,\ref{tomography} we present the emission-line ``tomography'', in \S\,\ref{discussion} we discuss our results, in \S\,\ref{feed} we estimate the mass outflow rate along the NLR and compare with the mass accretion rate and in \S\,\ref{conclusion} we present our conclusions.

\section{The Data}
\label{data}

The observations comprise three pointings obtained with the Gemini North telescope with the NIFS instrument \citep{mcgregor03} operating with the ALTAIR adaptive optics module, each covering a square field of view of$~\approx\,3.0\,\times\,3.0\,$arcsec$^2$. One of the NIFS fields was centred on the NGC\,4151 nucleus and the others were offset by$~\pm\,2.5\,$arcsec along position angle$~75\degr$, resulting in a total field of view of$~8.0\,\times\,3.0\,$arcsec$^2$, with the longest extent oriented  approximately along the radio jet \citep{pedlar93,mundell95,mundell03}. 

The spectra cover the Z, J, H, and K spectral bands at two-pixel resolving powers of $4990$, $6040$, $5290$, and $5290$, respectively. The angular resolution, as determined from the FWHM of the PSF, ranges from $0.12\pm0.02\,$arcsec in the H and K spectral bands to $0.16\pm0.02\,$arcsec in the Z and J bands. This corresponds to a spatial resolution at the galaxy of $8\pm1.3$ pc at H and K bands, and $10\pm1.3\,$pc in the Z and J bands.

More details about the data and the data reduction can be found in Paper\,I, as well as flux measurements of 61 emission lines from three characteristic locations along the NLR. In the present paper, we concentrate on the kinematics measured for the strongest emission lines of the ionised and molecular gas, namely:
\fsiii\,$\lambda\,0.9533\,\mu$m, \hei\,$\lambda\,1.0833\,\mu$m, \fpii\,$\lambda\,1.1886\,\mu$m, \pab\,$\lambda\,1.2822\,\mu$m, \ffeii\,$\lambda\,1.6440\,\mu$m and \hmol\,$\lambda\,2.1218\,\mu$m. We have also measured the kinematics for the coronal lines of  \fsix\,$\lambda\,1.2523\,\mu$m and \fcaviii\,$\lambda\,2.3220\,\mu$m. We first present and discuss the centroid velocity and velocity dispersion maps and then present the velocity tomography.

\section{Centroid velocity and velocity dispersion maps}
\label{results}

Centroid velocity and velocity dispersion~($\sigma$) maps were obtained by fitting Gaussians to the emission-line profiles. The velocities were obtained from the central wavelength of the Gaussians (after subtraction of the systemic velocity of $997$\kms, and applying a heliocentric correction of $+17$\kms) while the velocity dispersions were obtained from the full width at half maximum (hereafter FWHM) of the Gaussians as $\sigma=$FWHM$/2.355$.


In many locations along the NLR of NGC\,4151, the strongest ionised gas emission lines -- those of \fsiii\ and \hei\ -- show two kinematic components, with a mean velocity separation between them ranging from $200$ to $350$\kms. We were successful in fitting two Gaussian components to both these lines, although the \hei\ line has a broad component and a nuclear absorption component which introduces uncertainties in the fitting procedure for regions close to the nucleus. In a few regions, we observe double components also in \ffeii, \pab, and \brg. Elsewhere, these lines are broader at locations where double components are observed in \fsiii\ and \hei. This suggests that the  \ffeii, \pab, and \brg\   emission lines are also double. The lower signal-to-noise ratio (hereafter S/N) of these lines, the presence of a broad component for the \hi\ emission-lines close to the nucleus, and the fact that the \ffeii\ emission line appears to be intrinsically broader contribute to blend together two distinct kinematic components. 

Double components have been previously observed in the \foiii$~\lambda\,5007\,$\AA\ emission-line profiles by \citet{schulz90}, and \citet{crenshaw00a}. \citet{das05} using higher resolution (R$\,\sim\,9000$) spectra, showed that in some locations there are even three or four components. Our lower spectral resolution did not allow us to resolve more than two components.

\begin{figure*}
\includegraphics[scale=1]{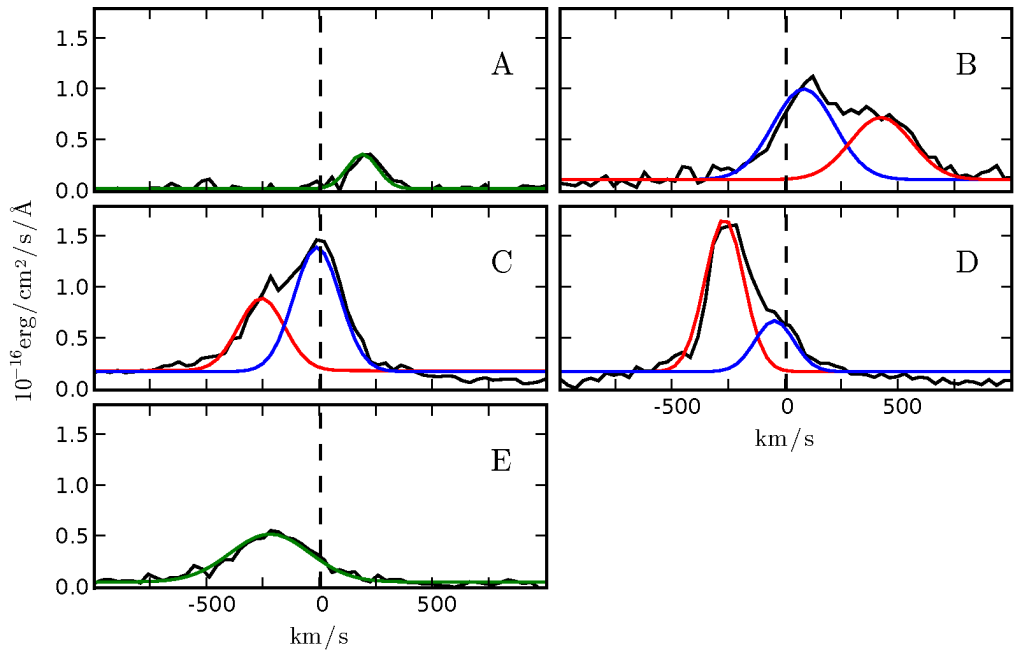}
\caption{Typical profiles and fits to the \fsiii\ emission line for the regions identified by letters A--E in Fig.~\ref{f-doublevel}.}
\label{f-profile}
\end{figure*}

\begin{figure*}
\includegraphics[scale=1]{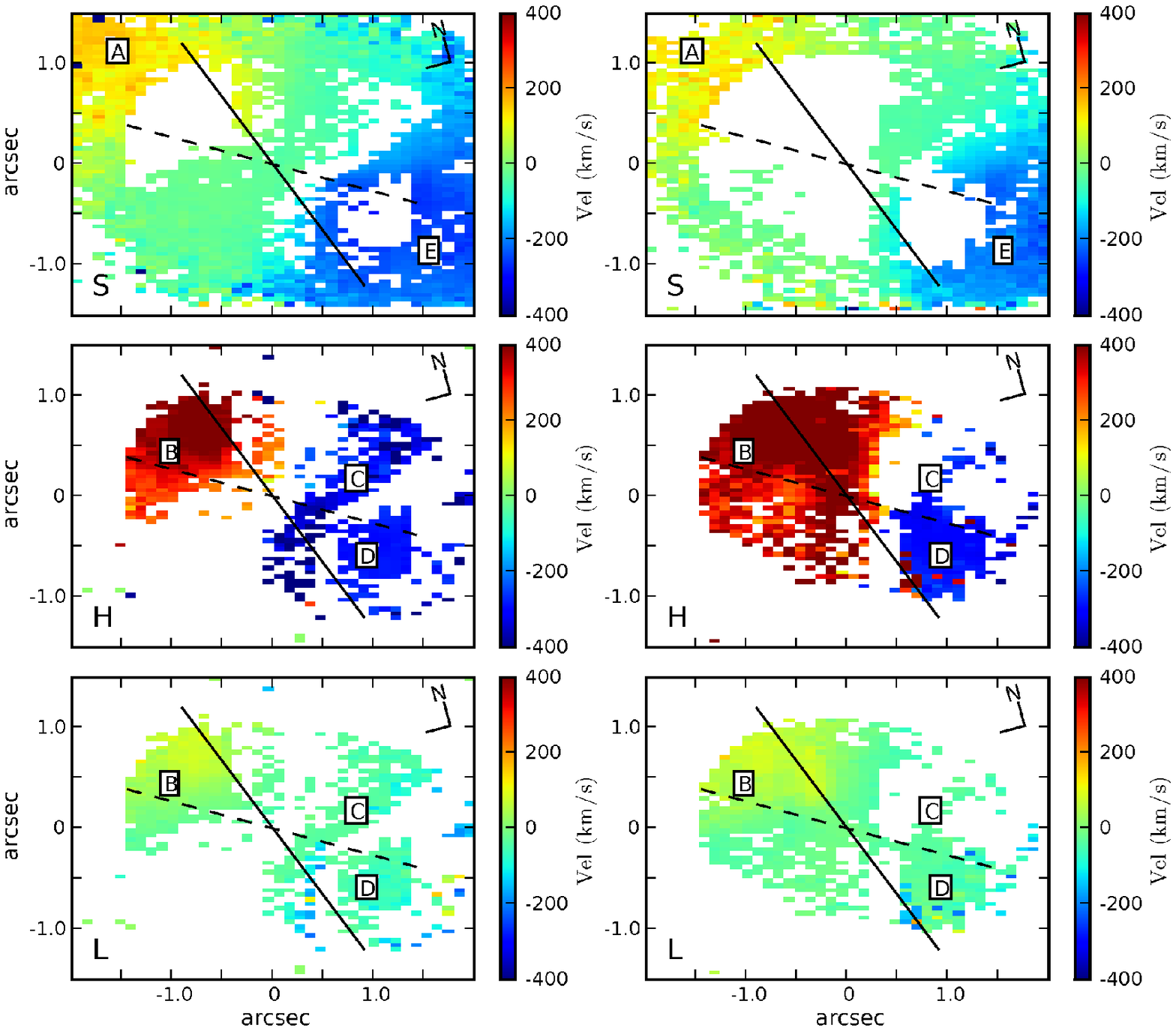}
\caption{Radial velocity maps for the single (top), and double components (middle and bottom panels) for \fsiii\ (left) and \hei\ (right). In the middle panel we show the high-velocity component, and in the bottom the low-velocity component. The continuous line shows the orientation of the galaxy major axis and the dashed line shows the orientation of the bi-cone. Emission-line profiles and fits typical of the regions marked with letters A--E are shown in Fig.~\ref{f-profile} for the \fsiii\ emission line.}
\label{f-doublevel}
\end{figure*}

Typical fits to the \fsiii\ emission-line profile are shown in Fig.~\ref{f-profile}, corresponding to locations (A, B, C, D and E) identified in the \fsiii\ radial velocity maps which are shown in Fig.~\ref{f-doublevel}. Because the S/N is not high enough to allow the fit of each component with the necessary number of free parameters, we had to constrain the width of the two Gaussian components to the same value. The implication of this assumption is that there will be some uncertainty in the gas kinematics derived in this way, mainly in the velocity dispersions, as discussed bellow. Although in the next section we derive the kinematics in terms of channel maps, with no assumptions, we decided to include the kinematics in terms of two components in order to be able to compare our results with those of previous studies, in which this was the method used to derive the gas kinematics. In addition, the separation in two components helps in the identification of the origin of the gas emission.

In the top panels of Fig.~\ref{f-doublevel} we show velocity maps in the emission lines of  \fsiii\ and \hei\ of regions with one kinematic component, after subtraction of the systemic velocity. In the middle and bottom panels we show the velocity maps of the regions with two kinematic components: the middle panel shows the high velocity component, corresponding to negative (blueshifted) velocities to the SW and positive (redshifted) velocities to the NE, while the bottom panel shows the low velocity component.

For the \fsiii\ emission, the lowest velocities are observed along the galaxy minor axis (PA\,$\approx-68\degr$) in the upper panel of Fig.\,\ref{f-doublevel} (single component), and in neighboring locations where two components were fit to the data. This low velocity component, essentially at zero velocity, probably originates in the galaxy disc, while the redshifted and blueshifted velocity components originate in the outflow along the bi-cone. The highest blueshifts are observed in the bottom part of the strip labelled ``C" in the middle left panel of Fig.\,\ref{f-doublevel}, while the highest redshifts are observed in the region labelled ``B".

\begin{figure*}
\centering
\includegraphics[scale=1]{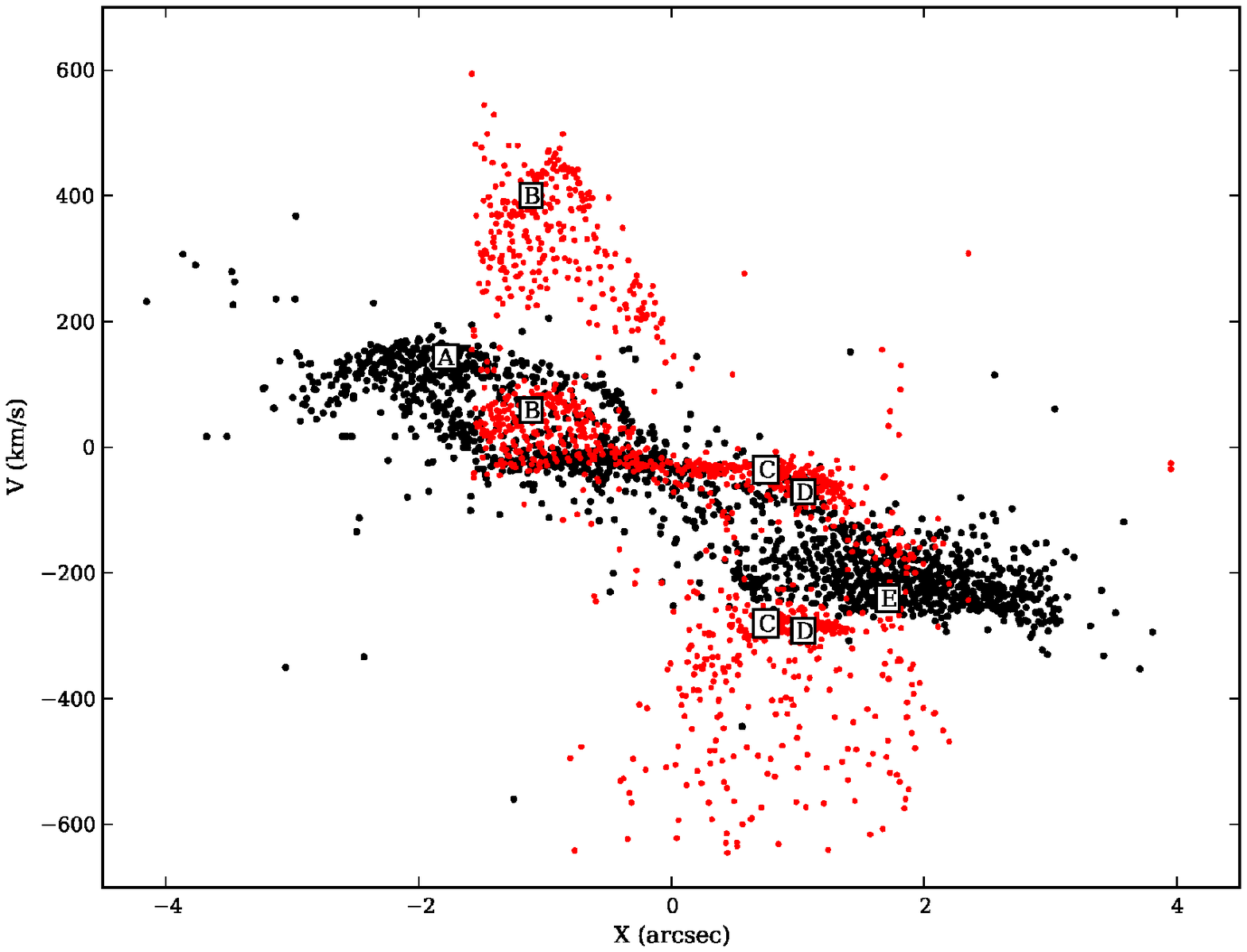}
\caption{Centroid velocity as a function of distance from the nucleus along PA$=60\degr$. Black circles represent the single-component fits, and red circles the two-component fits. Letters from A to E identify the components from the corresponding regions in Fig.\,\ref{f-doublevel}.}
\label{f-pv}
\end{figure*}

In order to better illustrate the magnitude of the velocity variations, and to investigate further the origin of the different gas components, we show in Fig.~\ref{f-pv}  the velocities as a function of distance from the nucleus along PA$=60\degr$. The black dots represent the single components and the red dots represent the double components, while the labels allow the comparison with Fig.\,\ref{f-doublevel}. Fig.~\ref{f-pv} suggests the presence of a $\approx$ zero velocity component observed between $\approx\,-1\farcs5$ and $\approx\,1\farcs5$ from the nucleus along PA$\approx\,60\degr$, whose origin is probably the galaxy plane, as pointed out above. Then there is  a sequence of blueshifts at $\approx -250$\kms\ to the SW (positive distances in Fig.\ref{f-pv}) which probably originates in the far side of the approaching cone (see Fig.\,\ref{f-geom} and discussion  in Sec.\ref{geom}). 
A similar sequence of redshifts at somewhat lower velocities, up to $\approx\,150$\kms\ to the NE  can be identified with the near side of the receding cone. The highest blueshifts to the SW and highest redshifts to the NE can then be identified with the near side of the approaching cone and the far side of the receding cone, respectively. These highest velocities thus originate in the walls of the bi-cone that are most inclined relative to the line-of-sight.


\begin{figure*}
\includegraphics[scale=1]{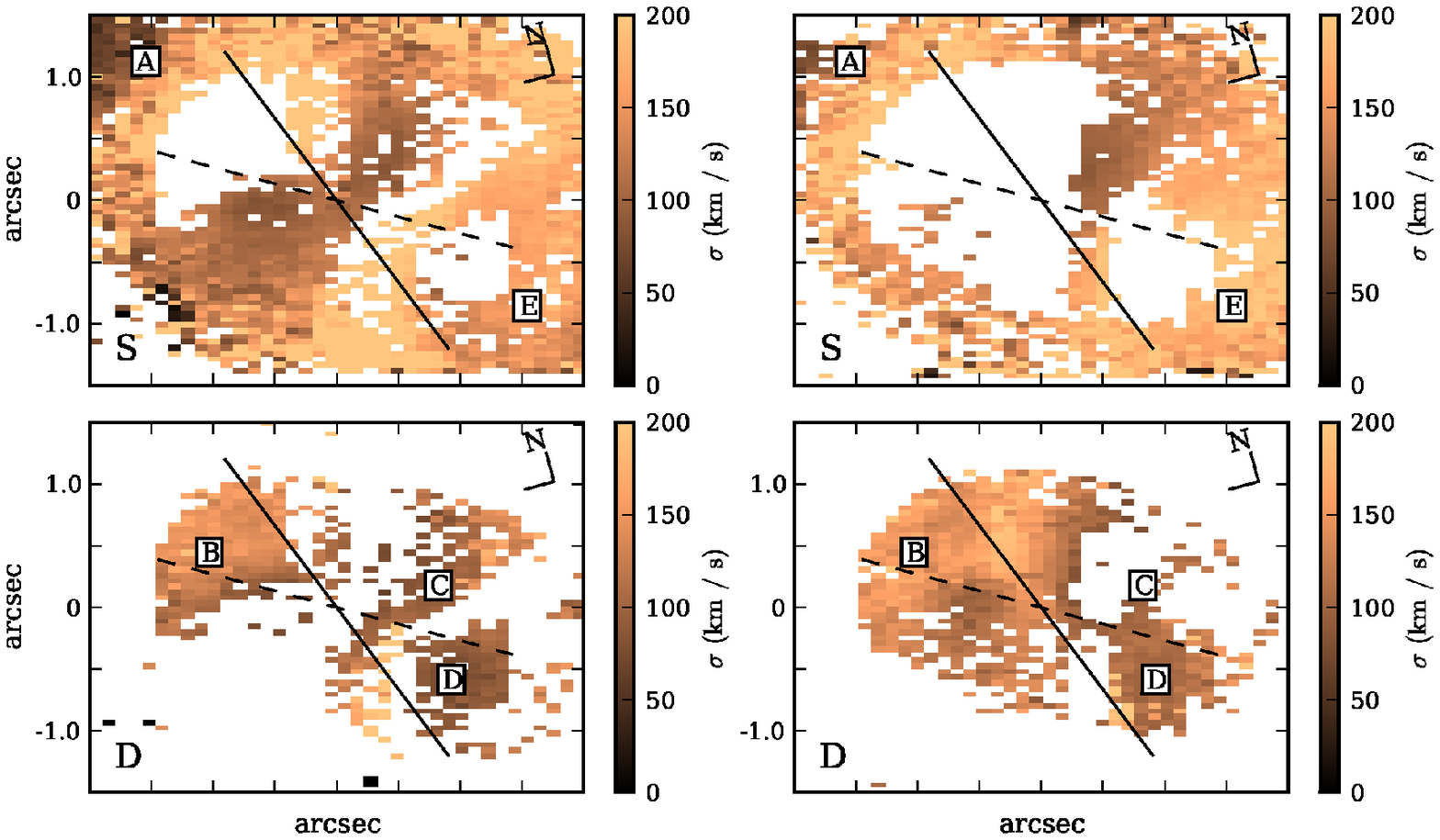}
\caption{Velocity dispersions obtained from the fits of the single (top) and double (bottom) components for \fsiii\ (left), and \hei\ (right). The continuous line shows the orientation of the galaxy major axis and the dashed line shows the orientation of the bi-cone.  Emission-line profiles and fittings typical of the regions marked with letters A, and E are shown on Fig.~\ref{f-profile} for the \fsiii\ emission line.}
\label{f-doublesig}
\end{figure*}

Velocity dispersion ($\sigma$) maps of the \fsiii\ and \hei\ emission lines for the single and two-component fits are shown in Fig.~\ref{f-doublesig}. We note again that the measured velocity dispersions are upper limits to the intrinisic gas velocity dispersion in regions where a single Gaussian component has been fit, possibly to multiple unresolved components. For example, in region E in Fig.~\ref{f-doublesig}, the velocity dispersion is $\ge$\,150\kms. But this large velocity dispersion is probably due to the blending of at least two components, as it surrounds a region with two components (region D). Similarly, the measured values are only indicative of the gas velocity dispersion in regions where two components have been fit, as  their widths have been constrained to the same value. This is the case of region E, where there are a component at zero velocity and a blueshifted component, which do not necessary have the same velocity dispersion. Nevertheless, these $\sigma$ values should not be too far off because they are of the order of the $\sigma$ value of the zero velocity single component along the minor axis of the galaxy.

 The lowest $\sigma$ values for the single components are $\sim$ 100\kms. These are seen mostly along the minor axis of the galaxy.
The low velocity dispersion suggests that the emitting gas in these regions is not disturbed and probably lies close to the galaxy plane. The velocity dispersion is larger in other regions. This could be due to turbulence caused by the outflow or to multiple velocity components. The $\sigma$ values for the double components are in some regions (e.g. region ``D") almost as low as the lowest values of the single components, suggesting that at least one of the components originates in the galaxy plane. Regions of higher $\sigma$ values in the two components are those labelled ``B" -- in the NE outflow -- and the bottom part of the strip ``C", where the highest blueshifts are seen in the centroid velocity maps, supporting an origin in the front side of the SW cone, as discussed above.

The centroid velocity distribution of \hmol\ emission lines is dominated by velocities close to systemic. This emission is concentrated along the minor axis of the galaxy, whose orientation approximately coincides with that of the galaxy bar. There are some redshifts to the NE mainly along the galaxy major axis and some blueshifts to the SW along both the galaxy major axis and the cone axis (see Fig.\,\ref{f-chmh2}). The velocity dispersions are mostly constant at $\approx$\,80\,\kms\, supporting an origin for this gas in the galaxy plane. The coronal line flux distributions are just barely resolved and their centroid velocities are also close to systemic, while their velocity dispersions are $\approx$\,100\kms.

\section{Emission-line scanning}
\label{tomography}

\begin{figure*}
\includegraphics[scale=0.90]{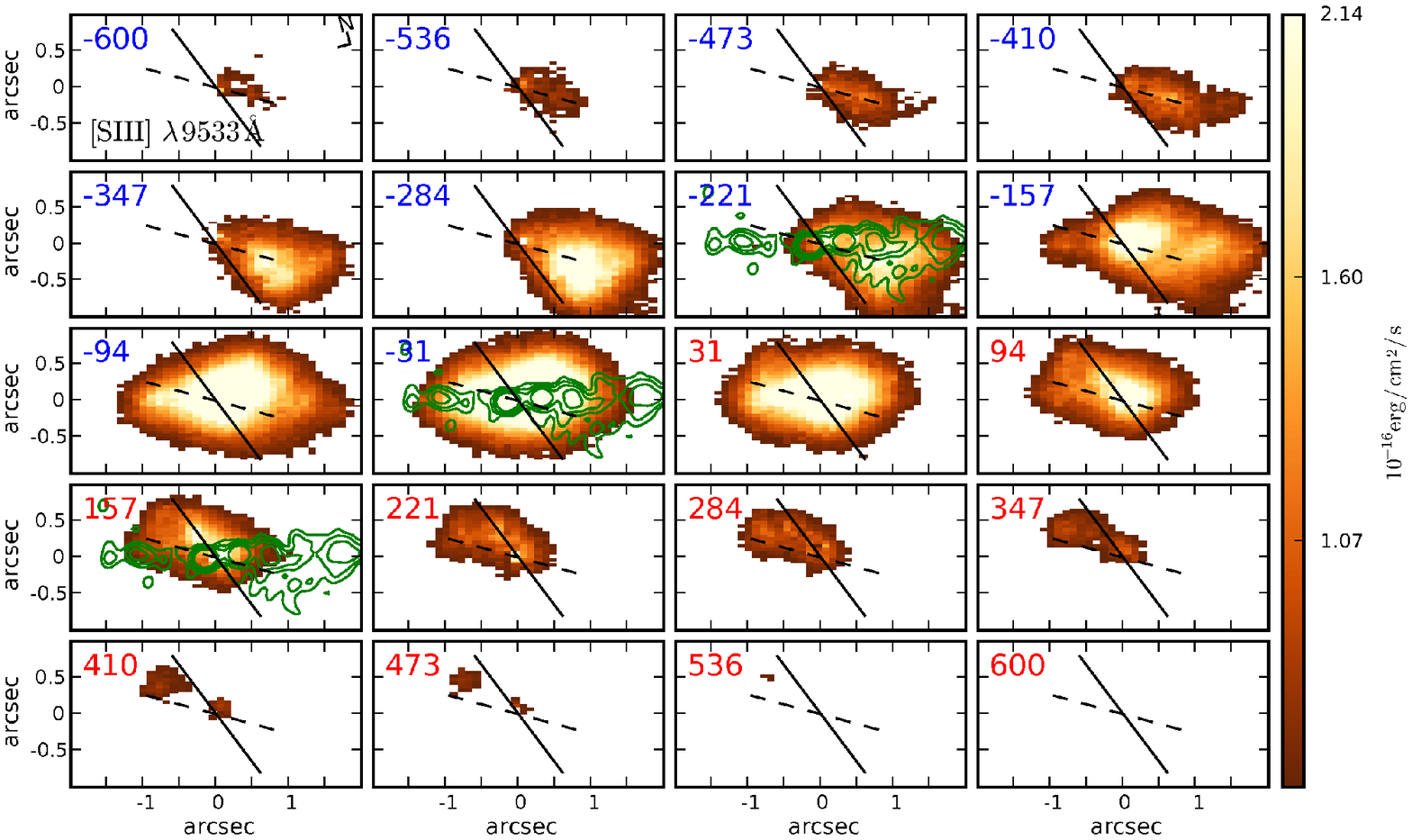}
\caption{Velocity channels obtained by integrating the flux within velocity bins of $63$\,\kms\ along the \fsiii\ emission-line profile. The numbers in the upper left corner of each panel are the central velocity of the bin, in \kms\ relative to systemic. The continuous line shows the orientation of the galaxy major axis and the dashed line shows the orientation of the bi-cone. Green contours are from the radio MERLIN image.}
\label{f-chmsiii}
\end{figure*}

\begin{figure*}
\includegraphics[scale=0.90]{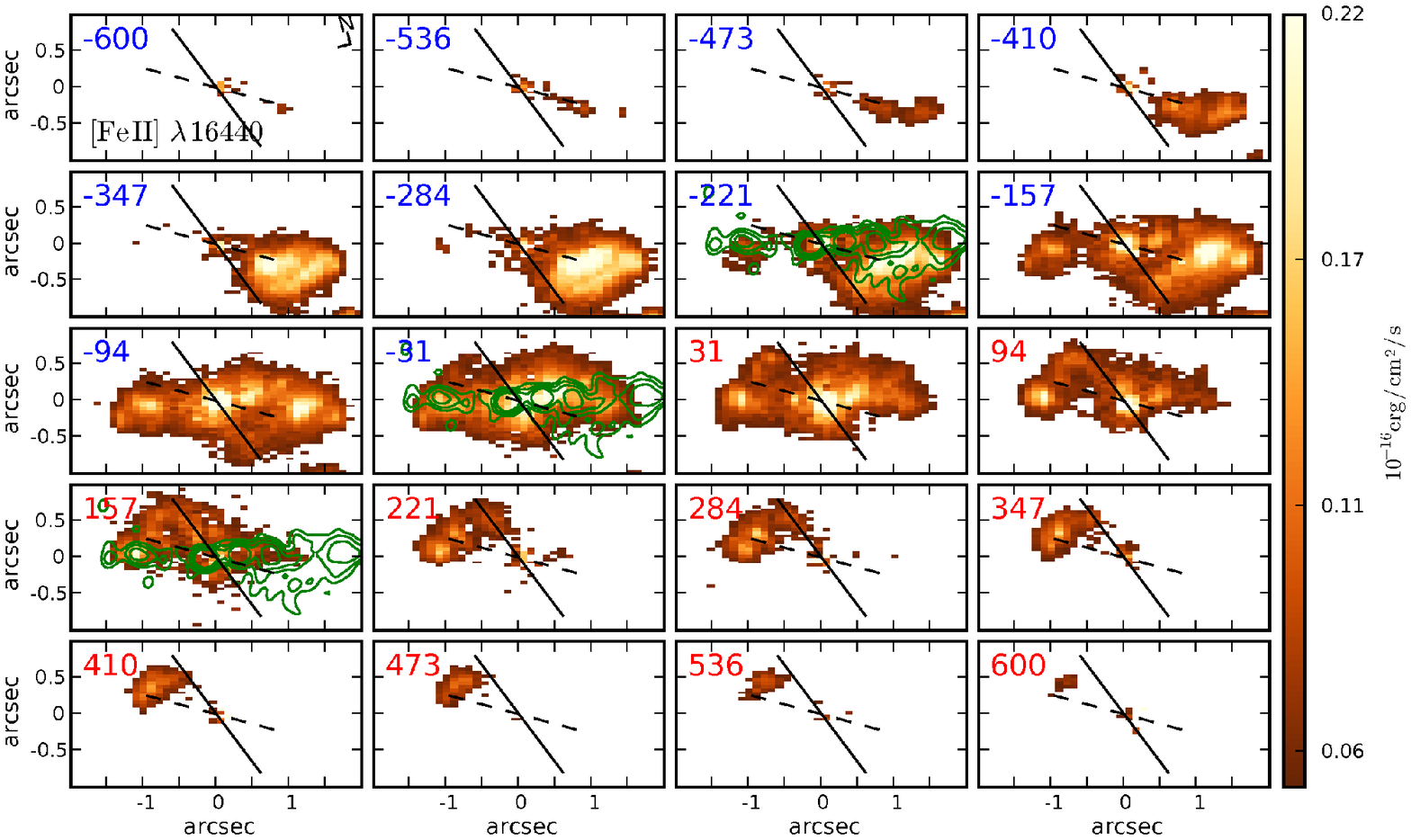}
\caption{Channel maps for the \ffeii\ emission line, as in Fig.~\ref{f-chmsiii}.}
\label{f-chmfeii}
\end{figure*}

\begin{figure*}
\includegraphics[scale=0.90]{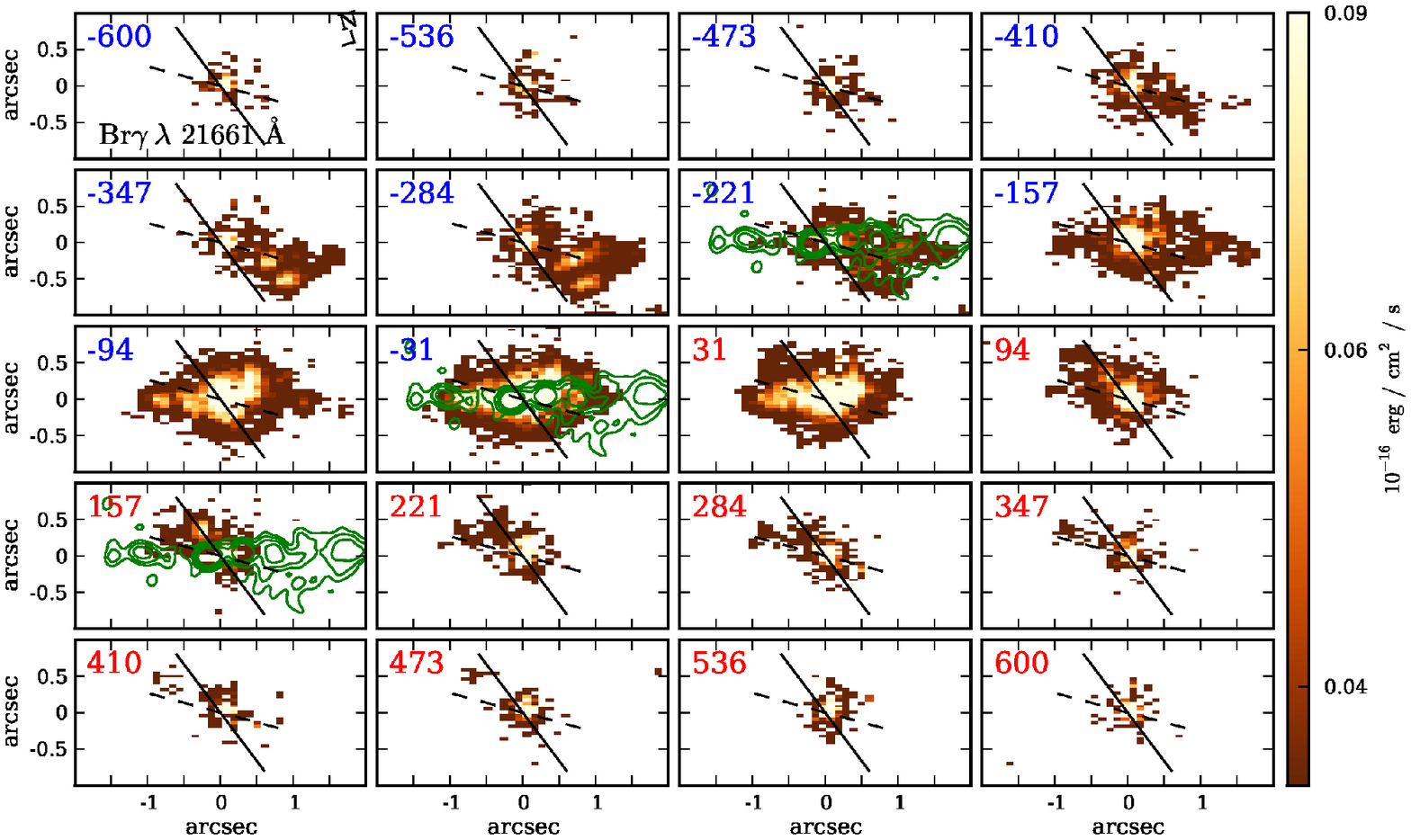}
\caption{Channel maps for the \brg\ emission line,  as in Fig.~\ref{f-chmsiii}.}
\label{f-chmbrg}
\end{figure*}

\begin{figure*}
\includegraphics[scale=0.90]{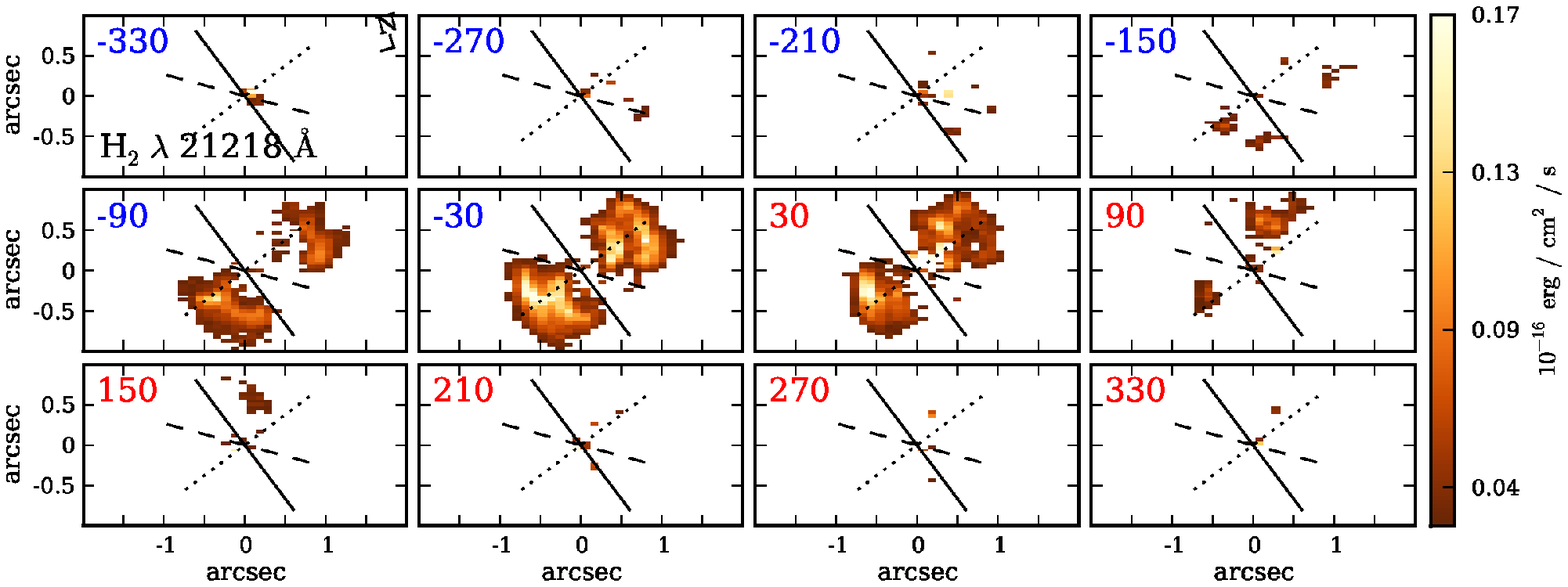}
\caption{Channel maps for the \hmol\ emission line, as in Fig.~\ref{f-chmsiii}. The dotted line shows the orientation of the galaxy minor axis.}
\label{f-chmh2}
\end{figure*}

We have also mapped the gas kinematics by integrating the fluxes within wavelength slices extracted along the emission-line profiles corresponding to velocity bins of $63$\kms. These ``channel maps" display the distributions of the gas emission at different velocities, which occur at different depths along the line of sight. In this sense, they can be described as velocities ``tomography" of the NLR using emission-line scanning. The channel maps are shown in Figs.~\ref{f-chmsiii}, \ref{f-chmfeii}, \ref{f-chmbrg} and \ref{f-chmh2} for the emission lines of \fsiii, \ffeii, \brg\ and \hmol, respectively. Channel maps in the other ionised gas emission lines, such as \hei\ and \pab, are similar to those in \fsiii\ and thus are not shown here. The lowest flux shown in the channel maps is $3$ times the root mean square variation of the surrounding sky flux. The bi-cone axis at PA$=60\degr$ as determined by \citet{das05} is represented by a dashed line in the figures, while the major axis of the galaxy at PA$=22\degr$ is represented by a continuous line. The contours of a radio MERLIN image (kindly made available by C. Mundell) have been overplotted on the central velocity bins (velocities close to systemic).

In the channel maps of the \fsiii\ emission line, shown in Fig.~\ref{f-chmsiii}, the most blueshifted gas at $\approx -600$\kms\ lies very close to the nucleus. For blueshifts of $\approx -400$\kms\ the emission extends further out, and for lower velocities broadens in the shape of a projected cone extending from the bi-cone axis towards the galaxy major axis. At the slice corresponding to $-221$\kms\ there seems to be a spatial association between the line emission and the radio structure. In particular, both of them broaden towards the major axis. At velocities close to zero, and up to redshifts $\approx 50$\kms, the emission is extended in all directions around the nucleus, evidencing the absence of a sharp cone apex. The largest elongation of this emission occurs along the horizontal axis, which is approximately the orientation of the radio axis. The overplotted radio contours show that, although there is not a detailed correspondence between the \fsiii\ and radio image, the brightest regions of the \fsiii\ image coincide with the two inner knots of the radio image. For redshifts larger than $100$\kms, the gas is observed mostly to the NE in the opposite direction to that of the blueshifted gas, showing also a similar flux distribution: for the lowest redshifts the emission is distributed in a fan structure extending from the major axis of the galaxy to the bi-cone axis, while for the intermediate and the highest redshifts the emission is more elongated and concentrated closer to the bi-cone axis to the NE. 

Channel maps for the \ffeii\ emission line are shown in Fig.~\ref{f-chmfeii}. In the high-velocity blueshifted channels, this emission seems to be more collimated along the bi-cone axis than for other emission lines. 
Some blueshifts from $\approx -221$\kms\ to zero velocity are observed also to the NE, unlike the slices in the other emission lines. The flux distributions for velocities close to zero and redshifted gas are similar to those in \fsiii, although the highest redshifts are as high as the highest blueshifts and are observed up to similar distances from the nucleus. In all velocity channels, there seems to be more structure in \ffeii\  than in \fsiii, in the form of knots of emission in closer association with the radio knots. This is particularly clear at the velocity bins from $\approx -221$ to $94$\kms: at $-221$\kms\ the strongest emission traces the ``flaring'' of the radio structure to SW towards the galaxy major axis; between velocities of $-45$ and $+45$\kms, the strongest emission approximately traces the two central radio knots and, for higher redshifts, there is an enhancement of the \ffeii\ emission at $\approx 1\,$arcsec NE, at the location of another radio knot.

Fig.\,\ref{f-chmbrg} shows the channel maps in the \brg\ emission line. These show similar morphologies to those seen in \fsiii, but with more structure. This can be attributed, at least in part, to the better spatial resolution achieved by the adaptive optics system in the K band relative to the Z band. 

The channel maps in the \hmol\ emission line, shown in Fig.~\ref{f-chmh2}, are completely different from those of the ionised gas, in support of a distinct origin for the molecular gas, as already pointed out in \citetalias{sb09}. Most of the emission occurs in the $\sim$ zero velocity channels. A small excess of blueshifts to the SW (from $\approx 100$\kms)  and similar redshifts to the NE suggest rotation around the minor axis. The flux distributions show a hollow core around the nucleus, and beyond this region the emission surrounds the minor axis of the galaxy in the shape of two arcs  extending up to $1\,$arcsec to NW and SE of the nucleus. A comparison with the channel maps in the ionised gas at the same velocity bins (around zero velocity) shows that the molecular gas emission avoids the innermost region of ionised gas, and the hollow structure can thus be explained by the presence of ionised gas surrounding the nucleus at zero velocity.


\section{Discussion}
\label{discussion}

\subsection{Velocity variations as a function of distance from the nucleus}

In the centroid velocity maps for the ionised gas emission (Fig.\,\ref{f-doublevel}), the lowest velocities are observed at the nucleus, while the highest are observed away from the nucleus. This is the case for NGC\,4151 and also for 6 other Seyfert galaxies for which we have recently obtained kinematic data with the IFU of the Gemini Multi-objetct Spectrograph (GMOS) \citep{barbosa09}.

The channel maps, on the other hand, show a different behavior: the highest velocities are usually observed at the nucleus \citep{barbosa09}. In the case of NGC\,4151, blueshifts of $\approx\, -600$\kms\ in \fsiii\ and \ffeii\  are observed at the nucleus and in a narrow region along the bi-cone axis extending from the nucleus almost to the outer edge of the NLR (Figs. \ref{f-chmsiii} and \ref{f-chmfeii}). Lower velocity blueshifted gas appears to fill the bi-cone and extend into the region of the radio jet near the systemic velocity. A similar structure is seen in redshifted gas with the emission at moderate redshifts appearing to fill the receding bi-cone. Emission at the highest redshifts of $> 400$\kms\ occurs at the nucleus and in a localised region $\sim$ 0$\farcs$9 to the NE of the nucleus that is offset from the bi-cone axis. This morphology differs from that of the highest velocity blueshifted gas and may simply reflect the clumpy nature of the NLR emission.

We attribute the difference between the kinematics probed by the centroid velocity and channel maps to the fact that the centroid velocity probes the brightest emission, while the channel maps also probe fainter emission (in the wings of the line profiles). In the vicinity of the nucleus the brightest component is the one originating in the  galactic disc, while away from the nucleus the brightest component is the outflowing one. As a consequence, the centroid velocities show an apparent increase in velocity outwards, that mimics acceleration along the NLR. In the channel maps we see high velocity gas at the nucleus, showing that the outflow does not leave the nucleus at zero velocity. We interpret this high velocity gas as an outflowing wind from the AGN or ambient gas more directly interacting with this wind. As this highest velocity gas moves away from the nucleus, it pushes and accelerates the gas from the disc.
The small amount of high velocity gas thus transfers momentum to a large reservoir of low velocity gas from the disc, causing the observed emission-line profiles and velocity structure along the bi-cone. This scenario is consistent with the one proposed by \citet{everett07} who, after verifying the failure of various wind models to reproduce the observed outflows in NGC\,4151, suggested that the interaction of an outflowing nuclear wind with the ISM could explain the observed velocity profile.

\subsection{Relation between the gas kinematics and the radio jet}

Some of the previous studies of the NLR of NGC\,4151 have suggested that there is no relation between the radio jet and the bi-conical outflow \citep{das05,crenshaw00a,hutchings99}. The distinct orientations of the two outflows do support distinct origins for them: the radio jet (oriented at PA$\approx$\,75$\degr$) originates in the funnel of the accretion disc, while the bi-conical outflow originates further out, probably still in the accretion disc. The disc should also be warped, in order to produce a wind with a distinct orientation (PA=60$\degr$) from that of the radio jet.
The channel maps of Figs.~\ref{f-chmsiii}, \ref{f-chmfeii} and \ref{f-chmbrg} confirm that  the flux distributions at high velocities show no relation with the radio jet.  But at velocities close to systemic the flux distributions become oriented along the radio axis instead of along the bi-cone. In addition, a comparison between the channel maps for velocities close to systemic and the contours of the radio map reveals line flux enhancements at the  locations of radio knots. Our proposed interpretation is that the radio jet has been launched close to the plane of the galaxy, and does interact with the circumnuclear gas pushing it out approximately along the plane and introducing the flux enhancements mostly due to gas density enhancement which are observed at velocities close to systemic. 

The observed flux enhancements could also be due to extra ionization by shock-induced X-rays in the NLR clouds interacting with the jet.  Recent Chandra X-ray observations of the NLR of NGC\,4151 \citet{wang09} have shown extended emission along the NLR with X-ray flux enhancements  in association with radio knots. They show that the radio-jet parameters are only consistent with the X-ray flux beeing due to thermal emission associated with the radio jet. 

The presence of shock heating is supported by the results of Paper\,I, namely the increase of the  [Fe{\sc\,ii}]\,1.257/[P{\sc\,ii}] line ratio to values larger than $\approx$\,2 in association with the radio knots. Values larger than $\approx$\,2 imply the presence of fast shocks to destroy dust grains and release the Fe. We also find, in Paper\,I, a high temperature (~16 000K) in the flux-enhanced regions, as derived from the [Fe{\sc\,ii}] line ratios, in agreement with this scenario. In the channel maps, the relation of the gas emission with the radio jet seems also to be strongest for \ffeii, as can be observed in Fig.~\ref{f-chmfeii}: at velocities from $\approx -200$ to $\approx +100$\kms, there is enhanced \ffeii\ emission at the locations of most radio knots. There seems also to be an association between a ``flared structure'' to the SW observed in the radio map with the line emission at velocities of up to $\approx -300$\kms. This suggests that the radio jet has been deflected towards the line-of-sight at this location. Some spatial association between optical and radio emission in the NLR of NGC\,4151 has also been observed by \citet{mundell03}. Comparing high resolution radio images to an HST \foiii\ image, they found a number of bright \foiii\ clouds that seem to bound a number of knots observed in the radio image, which they interpret are due to shocks of the radio jet with the circumnuclear ISM as the jet moves through the NLR and encounters gas clouds.

\subsection{Comparison with previous studies}
\label{geom}

\citet{crenshaw00b} studied the \foiii\ kinematics of the NLR of the prototypical Seyfert 2 galaxy, NGC\,1068, using a long-slit spectrum obtained at PA 202\degr\ with STIS on HST. They found a distribution of line centroid velocities along the slit that resembled a ``figure of eight'', with prominent high-velocity redshifted and blueshifted emission clumps and relatively few low-velocity clouds at intermediate distances from the nucleus. This distribution led them to infer a highly inclined conical geometry for the NLR of NGC\,1068, with the emission arising from clouds outflowing radially along the cone walls. The cloud outflow velocities increase approximately linearly with radial distance on both sides of the nucleus to $\sim \pm 500$\kms\ at distances of $\sim \pm 2\,$arcsec from the nucleus, and then decline back to zero by $\sim \pm 4\,$arcsec. \citet{crenshaw00b} were forced to hypothesise a gradual acceleration of the NLR clouds over $\sim \pm 140\,$pc followed by gradual deceleration to $\sim \pm 300\,$pc to account for this velocity structure. \citet{das06} used a more extensive set of STIS spectra to develop this model for NGC\,1068 further.

\begin{figure}
\centering
\mbox{\subfigure{\includegraphics{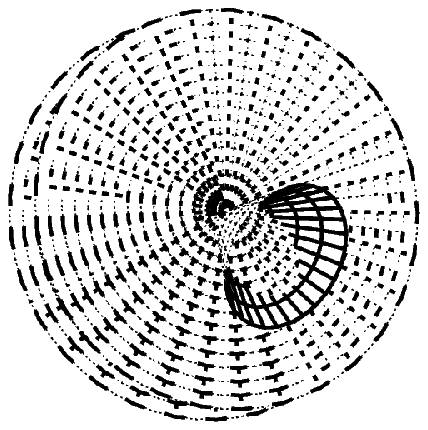}}
\subfigure{\includegraphics{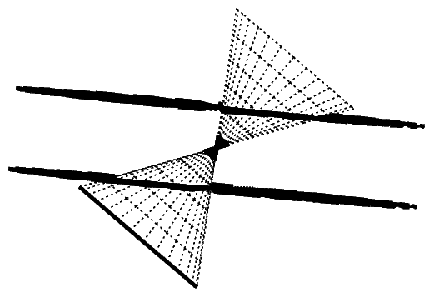}}}
\caption{Bicone geometry as proposed by \citet{das05}. In the left panel we show a representation of the bi-cone and galaxy disk at the observed orientation on the sky, while in the right panel the bi-cone is shown edge-on.}
\label{f-geom}
\end{figure}

\citet{crenshaw00a} applied the same model to STIS spectra of the NLR of NGC\,4151. The NLR of NGC\,4151 is far less inclined to the line of sight than NGC\,1068 -- \citet{crenshaw00a} infer $50\degr$ compared to $85\degr$ for NGC\,1068 -- so radial flows along the cone walls do not separate as cleanly into high-velocity redshifted and blueshifted structures. Consequently, the fit to the NGC\,4151 data is less compelling than for NGC\,1068. \citet{das05} revisited the NGC\,4151 NLR kinematics by fitting a similar model to spectra from five adjacent STIS slit positions. Their comprehensive analysis provides strong evidence that the NLR of NGC\,4151 is inclined by $45\degr$ to the line of sight, with a hollow bi-conical outflow occupying cone half-angles between $15\degr$ and $33\degr$. \citet{das05} conclude that the outflow along the walls of the bi-cone is best reproduced with a velocity law: $v \propto r$ valid for distances $r$ from the nucleus up to a turnover point at $r=96\,$pc ($=1\farcs48$), from where the gas starts to decelerate and reaches the systemic velocity at $400\,$pc ($=6\farcs15$).  The SW side of the bi-cone is tilted toward us, as illustrated in Fig.~\ref{f-geom}. In this geometry and orientation, we view the nucleus just outside the outer wall, at an angle of $12\degr$ with respect to the nearest part of the SW cone. This geometry and orientation has led to the idea that we see NGC\,4151 as a Seyfert 1 galaxy because the BLR can be viewed directly through a clumpy medium defining the cone walls.  The angle between the bi-cone axis and the normal to the galaxy disk is $36\degr$ \citep{das05}, and thus, for a total opening angle of the bi-cone of $66\degr$, one side of the cone exits the galactic disk steeply, with the outer wall making an angle of 87$\degr$ with the galaxy plane, while the other outer wall  leaves the plane at only 21$\degr$,  impacting a larger area of the disk (for a typical scale height for the disc of several tens of pc), as illustrated in Fig.\,\ref{f-geom}.

\subsubsection{Centroid velocity maps}
\label{centroid}

\citet{das05} argue that the intersection of the cone with the disc produces the observed geometry of the extended narrow-line region as shown by \citet{evans93}. If we now inspect our centroid velocity maps of Fig. \ref{f-doublevel} and Fig.\,\ref{f-pv} in light of the geometry proposed by \citet{das05}, we conclude that, in most regions where we see two velocity components, there is always one component at the systemic velocity plus an outflowing component in blueshift to the SW and in redshift to the NE. We attribute the systemic velocity component to emitting gas from the galaxy disc, in which case we are led to the conclusion that the AGN ionizes a region of the galaxy disc comparable in extent to that of the outflowing gas emission we detect along the bi-cone axis ($\approx$1\farcs5, or $\approx$100\,pc) and somewhat less extended perpendicular to it ($\approx$1\arcsec, or $\approx$65\,pc). 

Blueshifted components are observed to the SW at $\sim -250$\kms\ and at higher velocities of up to $-600$\,\kms\ (Fig.\,\ref{f-doublevel} for \fsiii\ and  Fig.\,\ref{f-pv}). We interpret the component with blueshifts $\sim -250$\kms\ at the position labelled  {\it D} in Fig.\,\ref{f-doublevel}, together with the single component surrounding it (region {\it E} in Fig.\,\ref{f-doublevel}), as originating in  the back surface of the SW cone. In the bottom part of the region we labelled {\it C} in Fig.\,\ref{f-doublevel}, we observe a few points with higher blueshifts of up to $-600$\,\kms\ (see also Fig.\ref{f-pv}). These components probably originate in the front part of the SW cone, which we see at a small angle ($12\degr$) and thus with a large velocity component along the line-of-sight. There is less emission from this front part of the SW cone than from the back, and our hypothesis to explain this is that  the cone walls are formed by gas entrained from the galaxy disk.  The front wall is thus likely to be comprised of more tenuous gas than the back wall, as there is less gas to be entrained at high angles to the galactic disk. 

Redshifted components are observed to the NE at somewhat lower velocities than to the SW (region labelled {\it B} in Fig.\,\ref{f-doublevel}). This can be seen more clearly in Fig.\,\ref{f-pv}: there is a sequence of redshifts of up to $\approx\,150$\kms, which can be attributed to the front of the NE cone and then some higher redshifts of up to 400--500\kms, which can be attributed to the back of the NE cone. In Fig.\,\ref{f-doublevel}, there is more emission from gas at the highest redshifts than emission from gas at the highest blueshifts. This result, together with the lower redshifts to the NE than blueshifts to the SW, suggest that the NE cone has its axis somewhat tilted towards the line of sight, instead of being aligned with the axis of the SW cone. This tilt would allow us to see more of the back wall of the NE cone and would result in lower observed velocities for both its front and back walls.

In the model proposed by  \citet{das05},  the NLR gas first accelerates up to $\approx$\,100\,pc from the nucleus and then decelerates, analogous to their successful model for NGC\,1068. While it is appealing in terms of AGN unification for NGC\,1068 and NGC\,4151 to be fit with a similar model, neither our centroid velocity maps of Figs.\,\ref{f-doublevel} and \ref{f-pv} nor the \foiii\ kinematic data for NGC\,4151 unambiguously point to this interpretation. The gradual acceleration over the inner $\sim100\,$pc and then deceleration required by the model remain unexplained. \citet{everett07} noted that radiative and magnetic acceleration from the region of the accretion disk both accelerate gas on length scales of the order of the launch radius, so winds driven by these mechanisms should reach their terminal velocities by $\sim 10\,$pc from the central source, not the $\sim 100\,$pc required by the model fits to the data. Thermal winds also fail in this regard \citep{everett07}. Dynamical models including various acceleration and deceleration mechanisms have been developed \citep{das07}. While these models fail to identify the acceleration mechanism, they do demonstrate that gravitational forces alone cannot produce the required deceleration, while drag forces from the interaction of NLR clouds with a diffuse, hot, X-ray-emitting ambient medium could be responsible.

The difficulty in identifying a suitable wind acceleration mechanism leads to the question of whether some of the NLR kinematics are due to gradual global wind acceleration or whether they result from the local interaction of an already accelerated wind with its ambient medium, as we have proposed above on the basis of the channel maps. This scenario has been proposed by \citet{everett07}, who realised that the interaction of an outflowing nuclear wind with the ISM could reproduce the apparent acceleration of the NLR following the works of \citet{matzner99} and \citet{lada96}. These authors showed that a protostellar hydromagnetic wind expanding into the surrounding gas produces a bipolar outflow with velocity proportional to the distance  to the star until a maximum radius beyond which the gas starts to decelerate.


\subsubsection{Position-Velocity diagrams}

\begin{figure*}
\centering
\includegraphics[angle=-90,scale=0.7]{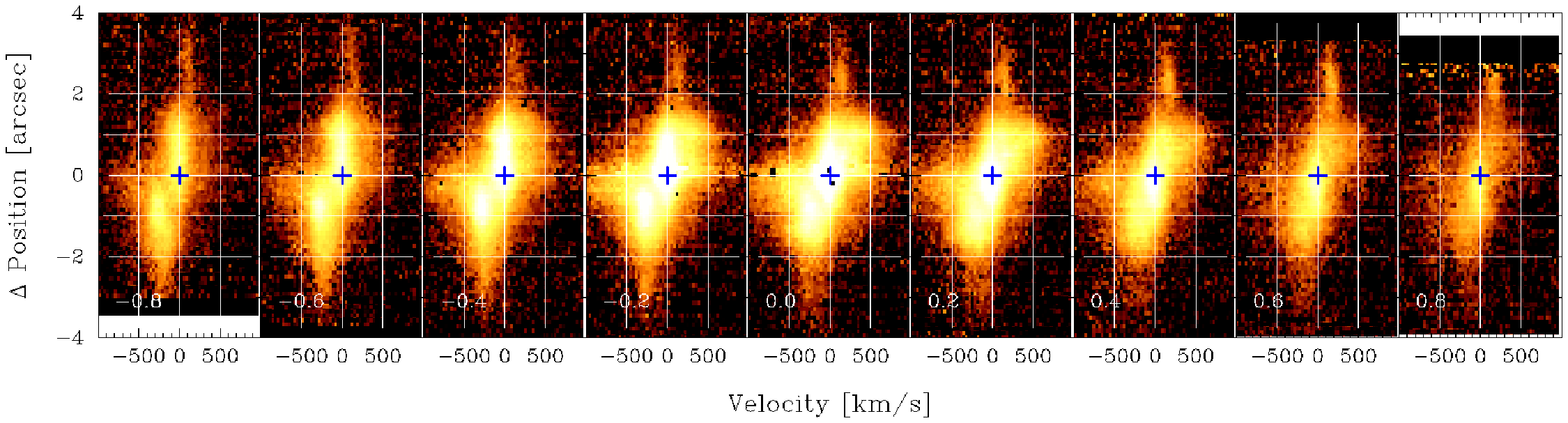}
\includegraphics[angle=-90,scale=0.7]{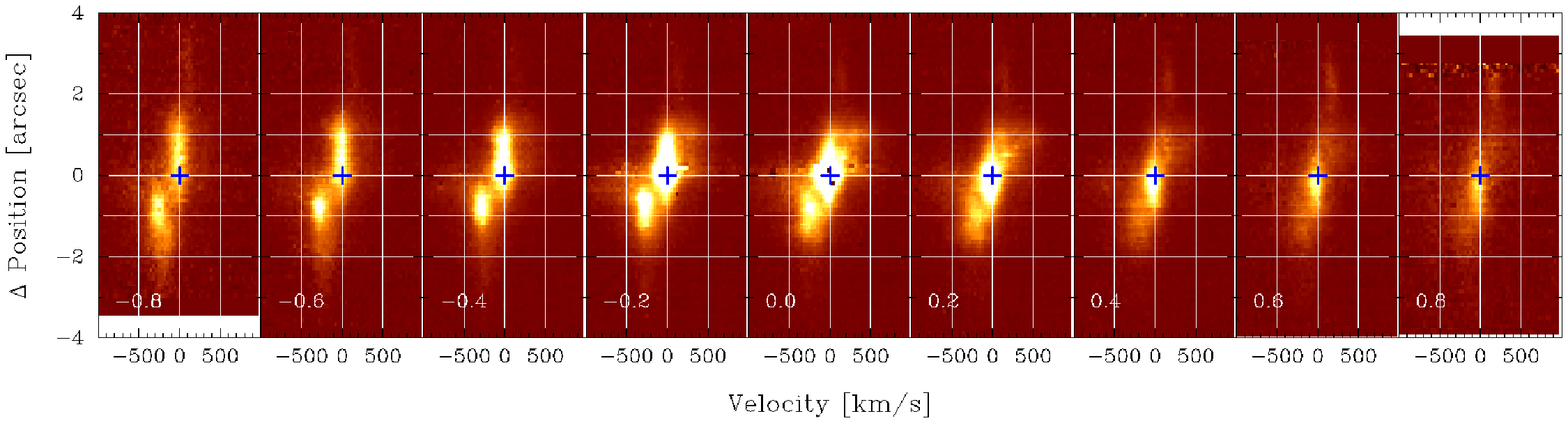}
\caption{Position-velocity diagrams of \fsiii\ 0.9533 \micron\
emission for 0.2\arcsec\ wide pseudo-slits oriented at PA=60\degr\
along the cone axis and offset perpendicular to the pseudo-slit by
-0.8\arcsec\ to +0.8\arcsec\ as indicated in each frame. An average
continuum image extracted between $\pm$(2000-2500) \kms\ has been
subtracted. Crosses mark the location of the nucleus in the central pseudo-slit and the ``equivalent" location for the other slits. Top: Line intensities are displayed on a logarithmic scale between $4.6\times10^{-16}$ and $4.6\times10^{-14}$ erg s$^{-1}$ cm$^{-2}$ \AA$^{-1}$ arcsec$^{-2}$. Bottom: Line intensities are displayed on a linear scale.}
\label{f-pv2siii}
\end{figure*}

\begin{figure*}
\centering
\includegraphics[angle=-90,scale=0.7]{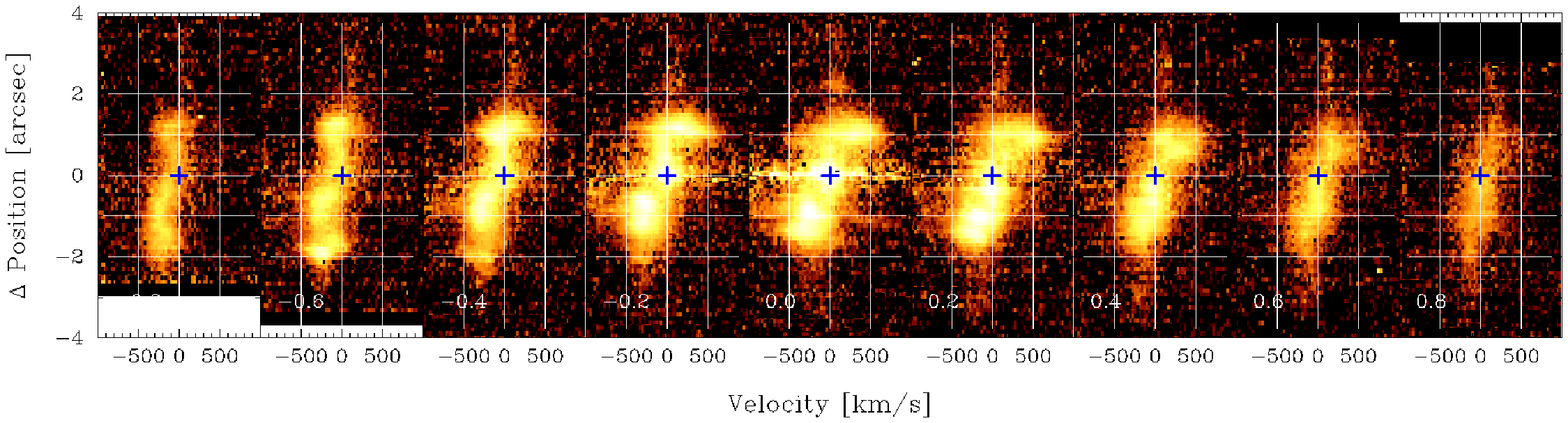}
\includegraphics[angle=-90,scale=0.7]{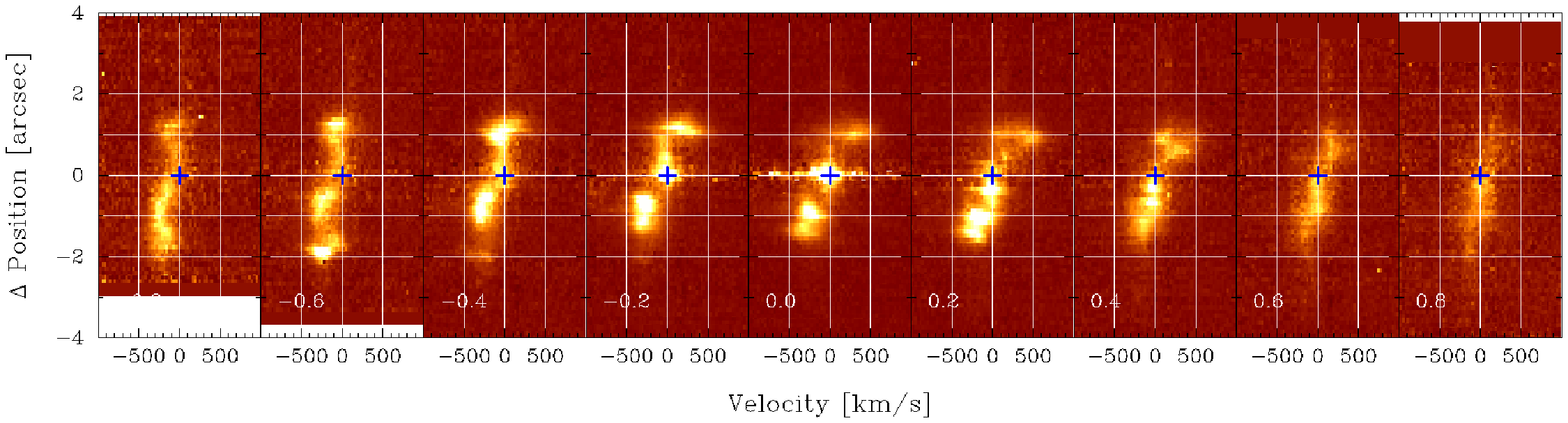}
\caption{Position-velocity diagrams of \ffeii\ 1.6440 \micron\
emission shown as in Fig.\ref{f-pv2siii}.}
\label{f-pvfeii}
\end{figure*}

In order to be able to compare our data with those of previous works (e.g. \citet{kaiser00}, \citet{das05}), we have built position-velocity diagrams (hereafter PV diagrams), which are shown in Fig.\, \ref{f-pv2siii} for the \fsiii\ emission line. These diagrams were obtained from the $Z$-band NIFS data cube by extracting $0.2\,$arcsec wide pseudo-slit spectra oriented at  PA$=60\degr$ (which is the orientation of the bi-cone axis), offset from the nucleus perpendicular to the pseudo-slit direction by between $-0\farcs8$ (along PA$=150\degr$) and $+0\farcs8$ (along PA$=-30\degr$) in steps of $0\farcs2$. The \fsiii\ diagrams are plotted both on a logarithmic scale (top of Fig.\, \ref{f-pv2siii}) to reveal the faint emission structure and on a linear scale (bottom) to show the regions with strongest emission. While qualitatively similar to previous long-slit  \foiii\ spectra \citep[e.g. Fig. 2 of][]{das05}, the \fsiii\ diagrams seem to show less emission at velocities exceeding $\pm 1000$\kms. In particular, high-velocity \foiii\ clouds A and B in \citet{kaiser00}  (near $-1400$\kms\ and $-1\farcs1$) are not apparent in our data. In part, this could be due to the poorer spatial resolution obtained with NIFS+ALTAIR at short near-infrared wavelengths.

As with the \foiii\ line, most of the \fsiii\ emission occurs at velocities below $\sim\,\pm\,700$\kms. The central diagrams of Fig.\, \ref{f-pv2siii}  can be compared directly with the \foiii\ slit spectrum shown in Fig.~2 of \citet{das05}. The \fsiii\ spectra are much less structured than the \foiii\ spectra, presumably due to poorer spatial resolution and consequent lower sensitivity to faint emission. Nevertheless, the two emission lines show a similar overall appearance along the central slit with blueshifted emission near $-250$\kms\ extending from $\sim -3\,$\arcsec\  to $\sim\,-0\farcs$3 from the nucleus towards PA$=-120\degr$ (SW, bottom of Fig.\,\ref{f-pv2siii}), emission at the systemic velocity near the nucleus, and some redshifted emission near $+150-200$\kms\ extending to $\sim 1\,$arcsec from the nucleus towards PA$=60\degr$ (NE, top of Fig.\,\ref{f-pv2siii}). The off-nucleus PV diagrams in Fig.~\ref{f-pv2siii} have a similar overall structure, but show that the $-250$\kms\ component is present in all negative offsets up to the slit at $-0\farcs$8. Of note is the fact that the velocity of this emission remains approximately constant over its full extent. The brightest $+200$\kms\ emission is only clearly apparent towards PA$=60\degr$ along the cone axis (i.e., in the upper part of the central panel of Fig.~\ref{f-pv2siii}). Some aditional emission from this component can be observed along the slits with positive offsets at fainter levels and at larger distances from the nucleus. The systemic velocity component is prominent in the upper section (NE) of pseudo-slits with negative offsets and in the lower sections (SW) of pseudo-slits having positive offsets, and originates in circumnuclear gas from the galaxy plane.

Fig.~\ref{f-pvfeii} shows the corresponding PV diagrams for the \ffeii\ 1.644 \micron\ line. The emission in the central pseudo-slit corresponds closely to the brightest \foiii\ emission in Fig.~2 of \citet{das05}, but the high-velocity extensions of the \foiii\ clumps are much less prominent in our \ffeii\ data. Although the major velocity components apparent in the \fsiii\ line are also apparent in \ffeii, the higher spatial resolution achieved at this longer wavelength helps to reveal sharper structures than in Fig.~\ref{f-pv2siii}. On the other hand, the structures seen in the \ffeii\ emission can also in part be attributed to enhancements associated with the radio jet, as observed in the channel maps. The component near $-250$\kms\ is comprised of several clumps and  extends to only $\sim -1\farcs$5 along the central pseudo-slit towards PA$=-120\degr$, but reaches $-2\,$arcsec at large negative, and possibly also positive, slit offsets -- i.e., in the extreme left and right panels of Fig.~\ref{f-pvfeii}. Clouds with velocities $\approx\,250$\kms\ are observed at $\sim +1\,$arcsec along the central slit and closer to the nucleus along the $0\farcs2$ and $0\farcs4$ offset slits. Emission  with velocities up to $+500$\kms\ is apparent at slit offsets of $-0\farcs2$ and $+0\farcs2$. The systemic-velocity component is restricted to the nucleus in the central slit, but extends to nearly $-1\,$arcsec along the $+0\farcs2$ to $+0\farcs6$ offset slits and  to nearly $+1\,$arcsec along the $-0\farcs2$ and $-0\farcs4$ offset slits. However, the systemic-velocity emission at negative slit offsets -- to the left in Fig.~\ref{f-pvfeii} -- is dominated by a further bright clump near $+1\,$arcsec -- approximately east of nucleus, in the vicinity of the receding radio lobe. Overall the impression from the \ffeii\ data is of roughly constant redshifted velocities towards PA$=60\degr$ and blueshifted velocities towards PA$=-120\degr$ with an additional systemic-velocity component originating in the galaxy disc and associated with the radio jet.

\subsection{Kinematic models for the NLR}

The role of the geometry and projection effects on the observations can best be evaluated via modelling. In this section, we compare our data with models with and without acceleration along the NLR.

\subsubsection{Das et al. Model}

\begin{figure*}
\centering
\includegraphics[angle=-90,scale=0.09]{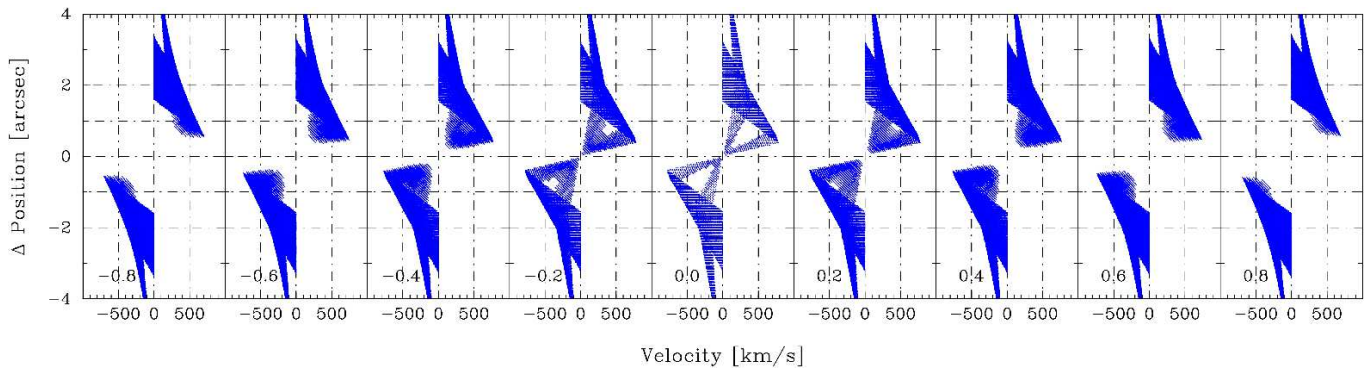}
\caption{Simulated position-velocity diagrams for the hollow cone with radial outflow model of \citet{das05}. The orientation and pseudo-slit geometry are as for Figs.~\ref{f-pv2siii} and \ref{f-pvfeii}.}
\label{f-daspv}
\end{figure*}

\begin{figure}
\centering
\includegraphics[angle=0,scale=0.45]{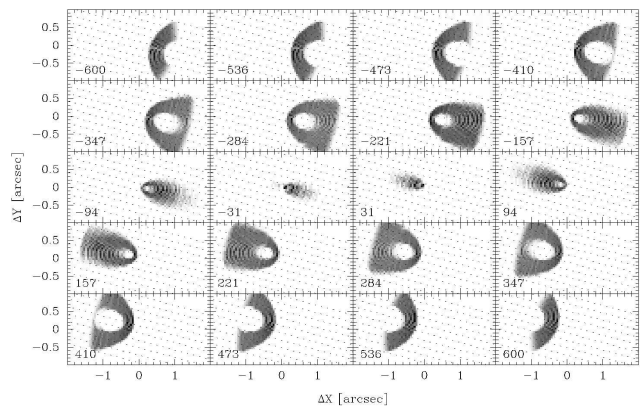}
\caption{Simulated channel maps for the hollow cone with radial outflow model of \citet{das05}. The orientation and velocity bins are as for Fig.\,\ref{f-chmsiii}.}
\label{f-daschm}
\end{figure}

We begin by testing the model proposed by \citet{das05} on the basis of the optical \foiii\ emission obtained with STIS, namely of an inclined hollow cone with radial outflow along the cone walls that is characterised by linear acceleration followed by linear deceleration \citep{crenshaw00a}. Our 3D kinematic data allow us to test this model against both channel maps and PV diagrams through multiple pseudo-slits in ways that were not available to previous authors. We begin by presenting both these views for the model parameters of \citet{das05} in Figs. ~\ref{f-daspv} and \ref{f-daschm}. As our data cover essentially the entire region of acceleration in the models, emission from decelerating gas beyond the velocity maximum has been suppressed in Fig~\ref{f-daschm}.

The \citet{das05} model predictions differ in several ways from the near-infrared emission-line data. First,  the high velocity gas just to the SW and NE of the nucleus predicted by the models (Fig.~\ref{f-daspv}) and seen in \foiii\ is less conspicuous in our data. It is seen in our data only in the log plots for \fsiii, where the faint emission gets enhanced (top panel of Fig.\,\ref{f-pv2siii}). We also see little evidence of acceleration and deceleration in the near-infrared data, which mostly show sequences at constant velocities from the nucleus all the way out. Even in the deep \fsiii\ PV diagrams, gas up to $3\,$arcsec from the nucleus in the lower half of these diagrams -- extending to PA$=-120\degr$ -- is still moving at blueshifted velocities of $\sim -250$\kms, with similar but smaller redshifted velocities in the upper half of the diagrams.


Another significant difference between the model and the near-infrared PV data is that the model predicts that pseudo-slits offset from the nucleus will show no emission near the slit centre -- Fig.~\ref{f-daspv}. This is the equatorial region away from the nucleus and between the two oppositely-directed cone ``lobes''. Most of the actual emission at off-nucleus slit positions arises in this region that is excluded by the model -- c.f., Fig.~\ref{f-pvfeii}. There is also emission at zero velocity -- mainly observed in the \fsiii\ PV diagram which is not included in the model.

Finally,  the model predicts that the channel maps should show ridges of emission perpendicular to the cone axis at high positive and negative velocities -- Fig.~\ref{f-daschm}. This morphology arises because outflowing material in the hollow cone essentially emits from an annular region perpendicular to the cone axis -- convolved with other orientation factors. In contrast, the channel maps in Figs.~\ref{f-chmsiii} and \ref{f-chmfeii}  show that the highest-velocity emission is confined to positions close to the cone axis, and is possibly extended along the cone axis towards the nucleus.



\subsubsection{Modified Das et al. Model}

\begin{figure*}
\centering
\includegraphics[angle=-90,scale=0.09]{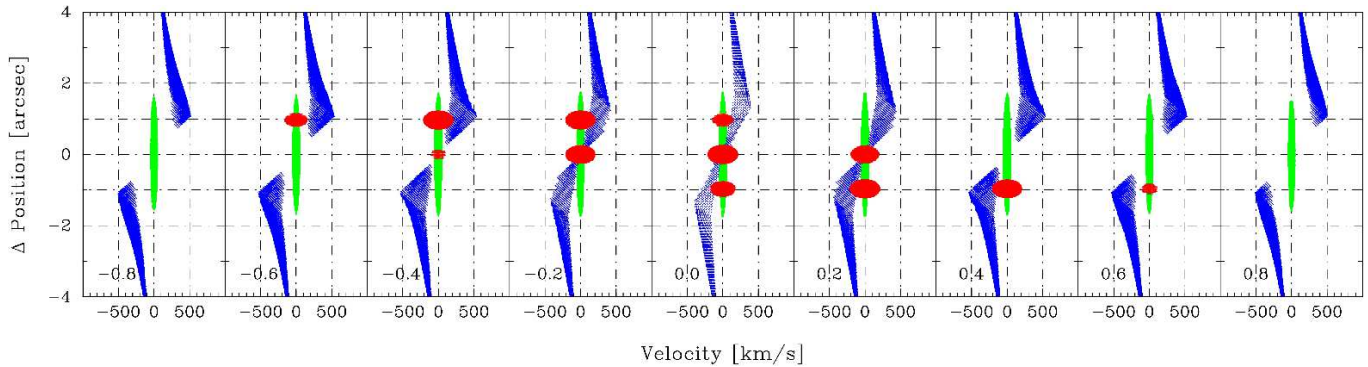}
\caption{Simulated position-velocity diagrams for the hollow cone with radial outflow model of \citet{das05} modified to include only the hemisphere of the cone with lowest inclination to the plane of the sky. The outflowing gas emission is shown in blue, an E-W radio jet component is shown in red, and an arbitrary galactic disk component is shown in green. The orientation and pseudo-slit geometry are as for Fig.~\ref{f-pv2siii}.\label{f-modpv}}
\end{figure*}

We now investigate modifications to the hollow-cone, radial-outflow model that attempt to address the above differences. The simplest modification is to hypothesise that the cone wall that is steeply inclined to the plane of the sky does not emit strongly in near-infrared lines. This could occur because this cone wall is more inclined to the galactic disk. As discussed in Sec. \ref{centroid}, under the hypothesis that the cone walls are formed by gas entrained from the galaxy disk, this wall is likely to be comprised of more tenuous gas, as there is less gas to be entrained at high angles to the galactic disk. This modification removes the high-velocity components in Fig.\,\ref{f-daspv}. To this, we then add an illustrative systemic-velocity component that is aligned with the radio jet and an arbitrary galactic disk component. The radio-jet component is modeled as a central nuclear component plus two clumps offset E-W from the nucleus by 1\arcsec. Each clump is assumed to be at the systemic velocity of the galaxy. The galaxy component has a Gaussian intensity profile and is assumed to be at the systemic velocity. The presence of this component is suggested both by the zero velocity component in the centroid velocity maps -- Fig.~\ref{f-doublevel} -- and by the channel maps, which show strong emission at zero velocities (Figs.~\ref{f-chmsiii} and \ref{f-chmfeii}). The resulting PV diagrams are shown in Fig.\,\ref{f-modpv}.

The PV diagrams in Fig.\,\ref{f-modpv} do not provide a perfect fit to the data (in particular to  the highest spatial resolution \ffeii\ data in Fig.\,\ref{f-pvfeii}). However, they serve to illustrate the following points: Firstly, the emission from the cone (blue regions in Fig.\,\ref{f-modpv}) is a better fit to the data  in the linear plots (showing the strongest emitting gas) if only the section of the cone wall closest to the plane of the sky, and hence closest to the galactic disk, is included. This may be due to interaction of the NLR gas in the cone wall with the galactic disk, possibly by the entrainment of disk material into the NLR outflow. 
Given that this material is moving in a direction closer to the plane of the sky, we are less sensitive to its true velocity so cannot make strong claims about the length scale on which it is accelerated. Fig.\,\ref{f-modpv} is drawn for the gradual acceleration and subsequent deceleration proposed by \citet{das05}. However, the data permit virtually any velocity law, including constant velocity outflow, on the length scales to which they are sensitive. The second point illustrated by Fig.\,\ref{f-modpv} is that emission from gas interaction with the radio jet is significant. This is modeled as nuclear and off-nuclear clumps, but in reality other emitting gas forms a bridge between the main clumps and the nucleus. This is seen in the channel maps shown in Fig.\,\ref{f-chmfeii}. Finally, the galaxy component at systemic velocity reproduces the data at PA=60$\degr$ for pseudo-slits at negative offsets and at PA$=-120$\degr for pseudo-slits at positive offsets, supporting the presence of a disc component, as concluded also from the centroid velocity maps of Fig.\,\ref{f-doublevel}.

\begin{figure}
\centering
 \includegraphics[angle=0,scale=0.45]{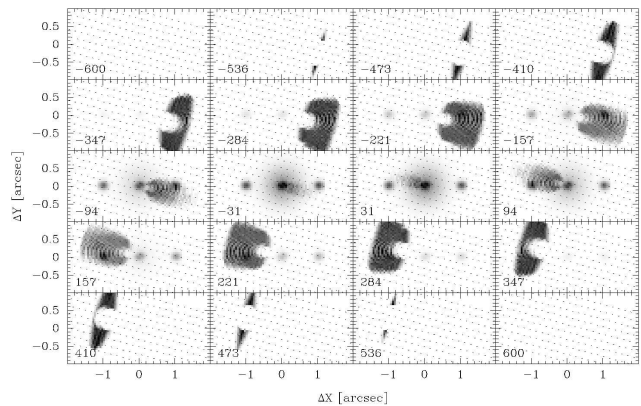} 
\caption{Simulated channel maps for the hollow cone with radial outflow model of \citet{das05} modified to include only the side of the cone with lowest inclination to the plane of the sky. The orientation and velocity bins are as for Fig.\,\ref{f-pv2siii}.\label{f-modchm}}
\end{figure}

Simulated channel maps are shown in Fig.\,\ref{f-modchm}. These only approximately reproduce the data presented in Fig.\,\ref{f-chmfeii}. In particular, two separated emission clumps aligned perpendicular to the cone axis are predicted from the cone walls at high-redshifted and high-blueshifted velocities while only a single clump near the cone axis is actually seen in the data.

\subsubsection{Constant Velocity Model}

\begin{figure*}
\centering
\includegraphics[angle=-90,scale=0.09]{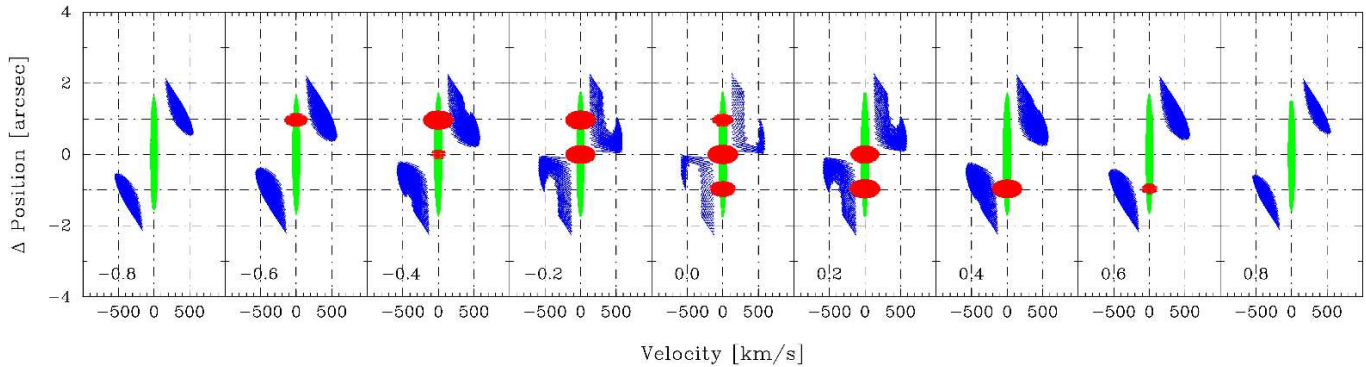}
\caption{Simulated position-velocity diagrams for the truncated hollow cone model with constant velocity $600$\kms\ outflow. The outflowing cone emission is shown in blue, an E-W radio jet component is shown in red, and the galactic disc component is shown in green (see text). The orientation and pseudo-slit geometry are as for Figs.~\ref{f-pv2siii} and \ref{f-pvfeii}.\label{f-modconstpv}}
\end{figure*}

Finally, we have considered a truncated hollow conical outflow at a constant velocity of $600$\kms, as we could not see acceleration or deceleration in our data. As in the previous model, we add the radio jet and galactic disk component. The resulting position-velocity diagrams are shown in Fig.~\ref{f-modconstpv}.

The fit to the observed constant velocity sequence at $\sim -250$\kms\ for \fsiii\ (Fig.\,\ref{f-pv2siii}) along the slit passing through the nucleus and the ones with negative offsets are improved by this model. The disc  component at zero velocity reproduces emission observed at the nucleus and along PA$=60\deg$ (positive position in the panels of Fig.~\ref{f-pv2siii}) and PA$=150\deg$ (to the left in Fig.~\ref{f-pv2siii}), as well as less extended emission along PA$=240\deg$ (negative positions in Fig.~\ref{f-pv2siii}) and PA$=330\deg$ (to the right in Fig.~\ref{f-pv2siii}). The high velocity sequences predicted by the models -- although not conspicuous in the linear PV diagrams of \fsiii\ (bottom panel of Fig.\,\ref{f-pv2siii}) -- are observable in the log PV diagrams (upper panel of Fig.\,\ref{f-pv2siii}), due to its faint emission. These high velocity sequences and are due to the front wall of the approaching cone and far wall of the receeding cone. We point out that this short high velocity sequence as predicted by the constant velocity model seems to be adequate to reproduce also the high velocity gas observed in  \foiii\ (Fig. 2 of \citet{das05}).

Nevertheless, this model did not improve the fit for pseudo-slits offset by more than $0\farcs4$ from the nucleus, which still do not show emission near the slit centre in the model.  In addition, these pseudo-slits show velocity sequences that decrease with increasing distance from the nucleus. 
In order to improve the fit, new geometries should be considered. For example, a flow that is not conical, but in the form of a bowl as observed in planetary nebulae and in recent NIFS observations of NGC\,1068 \citep{McGregor10}.

\begin{figure}
\centering
\includegraphics[angle=0,scale=0.45]{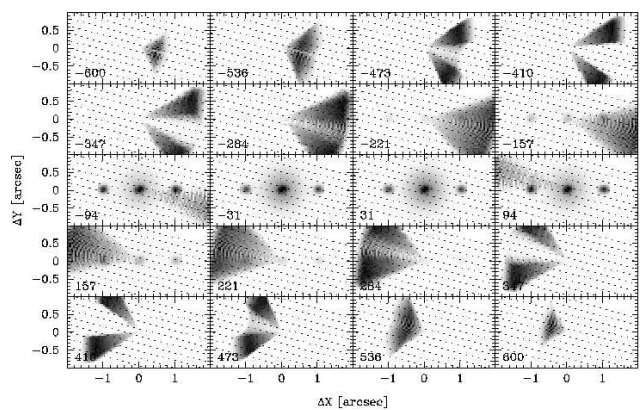}
\caption{Simulated channel maps for the truncated hollow cone model with radial outflow at constant velocity, as in Fig.~\ref{f-modpv} (see text). The orientation and velocity bins are as for Fig.\,\ref{f-pv2siii}.
\label{f-modconstchm}}
\end{figure}

Simulated channel maps are shown in Fig. \ref{f-modconstchm}. These maps are an improvement relative to the previous model for velocity channels between $-300$\kms and $+300$\kms, but show double structures for velocities between $300$\kms and $500$\kms, which are not seen in the data. For the highest velocities, a small triangular structure is seen, while the data show only clumps near the cone axis. This discrepancy can also be attributed to possible deviations from a bi-conical geometry and to very weak emission from the  wall of the bi-cone that is most inclined.

\subsubsection{Integrated Intensity}

\begin{figure}
\centering
\includegraphics[angle=-90,scale=0.32]{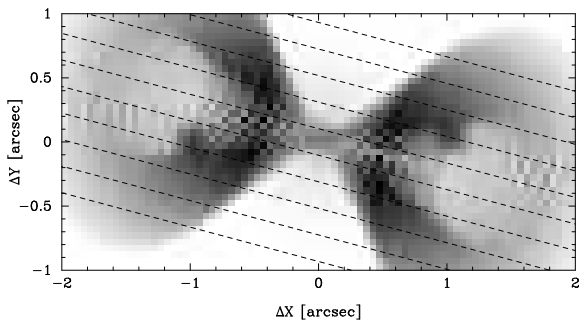}
\caption{Simulated integrated intensity map for the modified hollow cone outflow model of \citet{das05} (see text), to be compared to Fig.~\ref{f:FeII_Int}. \label{f:Mod_Int}}
\end{figure}

\begin{figure}
\centering
\includegraphics[angle=-90,scale=0.32]{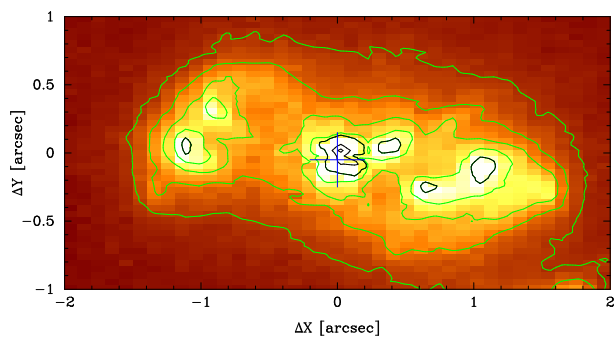}
\caption{Integrated \ffeii\ 1.644 \micron\ intensity formed by summing the velocity slices in Fig.~\ref{f-chmfeii}.\label{f:FeII_Int}}
\end{figure}

The modelled integrated emission-line map is similar for the three models discussed above. We show in Fig.~\ref{f:Mod_Int} the map obtained for the modified \citet{das05} model. A comparison with Fig.~\ref{f:FeII_Int} shows that it provides a reasonable reproduction of the data, taking into account that one should add the contribution of gas from the galaxy plane to the modelled emission.

\subsubsection{The origin of the highest velocity gas seen in \foiii}

\citet{cecil02} identified blueshifted \foiii\ emission with velocities up to $3200$\kms\ in the NGC\,1068 NLR $70-150\,$pc NE of the nucleus. This is the region attributed to the highest radial outflow velocities in the gradual acceleration model. Many of these knots are linked kinematically to more massive and slower-moving clumps located closer to the nucleus. \citet{cecil02} identify these knots as ablation flows from disintegrating molecular clouds that are being photoionised and radiatively accelerated by the AGN. \citet{hutchings99} detected faint, high-velocity \foiii\  emission with radial velocities up to $1400$\kms\ in STIS slitless spectra of NGC\,4151. The high-velocity emission was shown to arise in high-excitation ionised gas, and has velocity dispersions in excess of $130$\kms\ \citep{kaiser00}, which is also consistent with an origin in an ablation flow.

Near-infrared \fsiii\ emission arises in conditions similar to the optical \foiii\ emission, and can thus be formed in tenuous fully ionised gas experiencing a high local ionisation parameter. But then, why do we see less \fsiii\ than \foiii\ emission? Either we lack the sensitivity to detect faint high-velocity emission, or this emission is not present.

\begin{figure}
\centering
\includegraphics[angle=-90,scale=0.35]{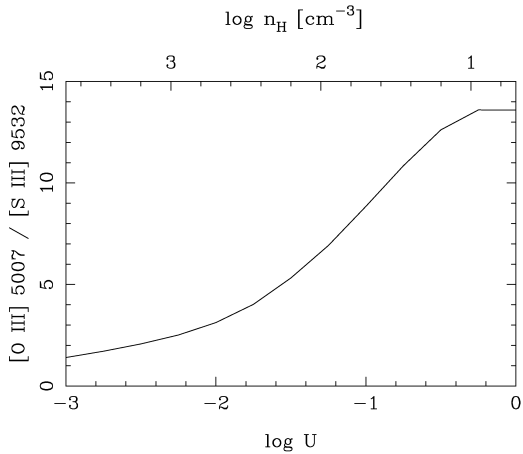}
\caption{\foiii/\fsiii\ line ratios for different ionisation parameters or densities with a fixed radiation field that corresponds to an ionisation parameter of $\log U=-2.0$ at a NLR cloud with a density of $500\,$cm$^{-3}$.\label{f-ratio}}
\end{figure}

In order to investigate this, version 8.0 of the photoionisation code Cloudy, last described by \citet{ferland98}, was used to calculate the ratio of \foiii\  to \fsiii\  for NLR clouds with a range of densities. The ionisation parameter was set to $\log U=-2.0$ for a gas density of $500\,$cm$^{-3}$ and then varied inversely with the gas density to simulate clouds with different densities being illuminated by the same radiation field. Such a situation would occur if the low-velocity \foiii\ is emitted at the front face of a dense NLR cloud and the high-velocity \foiii\ emission arises in an associated low-density ablation flow. The resulting emission-line ratios are plotted in Fig. \ref{f-ratio} where it can be seen that \foiii/\fsiii\ increases from $\sim 3$ at densities typical of NLR clouds to $\sim 13$ at densities $< 10$ cm$^{-3}$ that might be typical of an ablation flow. The values at high densities are typical of the integrated line ratios observed in Seyfert galaxies NLRs \citep[e.g.,][]{sb95}. The strength of \fsiii\ falls sharply to lower densities while the strength of \foiii\ is largely unchanged. The reason for this dependence on ionisation parameter is that sulphur exists predominantly as \fsiii\ at typical NLR densities, but as \fsiv\ and higher species in more highly excited low-density regions. In contrast, the higher ionisation potential of \foiii\ (54.9 eV compared to 35.1 eV for \fsiii\ and 47.2 eV for \fsiv) means that \foiii\  survives better in highly excited regions. The lower intensity of \fsiii\ in low-density irradiated regions makes it more difficult to detect a low-density ablation flow in \fsiii\ emission than in \foiii\ emission. This combined with the lower Strehl ratio and consequent poorer image quality achieved by adaptive optics systems at short near-infrared wavelengths probably account for our inability to detect more high-velocity gas in NGC\,4151.

There is little evidence in our near-infrared data for the gradual acceleration followed by gradual deceleration proposed by \citet{das05} based on \foiii\ kinematics. No viable physical mechanism has been identified for such a gradual acceleration. A possible explanation is that the high-velocity \foiii\ emission does not represent the bulk motion of the outflow. Rather, this could be low-density material ablated from NLR clouds that we observe outflowing at $\approx$\,250\,\kms\ and are accelerated much closer to the AGN. This is consistent with the low surface brightness of the high-velocity, low-density emission in \fsiii\ and with the prominence of near-infrared emission formed on the sides of the outflow closer to high-density gas in the galactic disk.

\section{Feeding vs. Feedback}
\label{feed}

While the kinematics of the \fsiii, \hei, \hi, and \ffeii\ emitting gas is dominated by outflows along the bi-cone, the \hmol\ radial velocity field in Fig.~\ref{f-chmh2} is dominated by velocities close to zero around the galaxy minor axis.  Although the velocity range is small, it can be observed that, from the blueshifted to the redshifted channels, the flux distribution moves around the minor axis of the galaxy, from the SW to the NE, as expected for gas rotating in the galaxy plane. Although we see a kinematic component at systemic velocity also in the ionised gas, we do not see the outflowing component in \hmol. The velocity dispersion in the \hmol\ gas is also the smallest among the emission-line profiles.

Distinct kinematics for the \hmol\ emitting gas relative to the emission lines of ionised gas has also been observed in previous studies of other objects \citep{sb99,ardila05,riffel06}. The \hmol\ emission lines usually have small velocity dispersions and their kinematics seem to be dominated by rotation in the galaxy plane, while the other emission lines have larger velocity dispersions and other velocity components besides rotation, as observed also in NGC\,4151. In Paper\,I we have argued that the flux distributions, physical properties and location suggest that the \hmol\ emission gas traces the AGN feeding, while the ionised gas traces its feedback.  Although we do not see the \hmol\ emitting gas actually flowing down into the nucleus, it does present a distinct kinematic signature relative to the ionised gas. The inflow may be occurring further in, in non-emitting colder molecular gas. The hot molecular gas is estimated to be a fraction of only $10^{-7}$ to $10^{-5}$ of the total mass of molecular gas \citep{dale05} in AGN. In the case of NGC\,4151, for which we obtained a mass of hot \hmol\ of $\approx 240\,$M$_\odot$ \citepalias{sb09}, we can thus estimate a total molecular gas mass of$~\sim\,10^8\,$M$_\odot$. Such a large molecular gas reservoir around the nucleus of active galaxies is supported by radio observations of the central region of active galaxies in general, by, for example, the ``NUGA'' group, which report molecular gas masses in the range $10^7$--$10^9\,{\rm M_\odot}$ in the inner few hundred parsecs of active galaxies \citep[e.g.][]{boone07,gb05,krips07}. If such a reservoir of molecular gas has accumulated by a flow that was the trigger of the present nuclear activity, it can be concluded that a steady flow would have to bring gas inwards at a rate  in the range $0.1$--$10\,$M$_\odot\,$yr$^{-1}$. 
In NGC\,4151, \citet{mundell99} have measured inflows in radio observations of \hi\ that occur along narrow channels interpreted as shocks along the leading edges of the large scale bar (at PA$\,=\,130\degr$). These inflows seem to end at the inner part of the bar in the nuclear region. The orientation of the flow is approximately perpendicular to the ``arcs'' seen in the \hmol\ flux distribution -- see Fig. \ref{f-chmh2} -- and one possibility is that the radio observations are tracing the gas inflows that accumulate in the nuclear region and contribute to build up the molecular gas, part of which we observe due to excitation of the \hmol\ molecule by X-rays emitted by the nucleus.

The large amounts of molecular gas observed around the nuclei of active galaxies can only lead  to the conclusion that  inflows must be occurring \citep{gb05}. Nevertheless, the actual kinematic observation of inflowing gas in active galaxies is difficult to obtain. This is due to the fact that the emission of the cold, inflowing gas is much weaker than the emission of the hotter, outflowing gas. In \citet{fathi06} and \citet{sb07} we were successful in mapping the inflow in low-excitation ionised gas because the targets were low-activity AGN -- the LINERs NGC\,1097 and NGC\,6951 -- for which the nuclear radiation and outflows are much weaker than in Seyfert galaxies, thus allowing the observation of inflows in the disc in ionised gas. We are probably seeing just the ionised ``skin'' of a flow of cold gas on its way to feed the AGN. We were also successful in mapping an inflow observed in hot molecular gas in the galaxy NGC\,4051 \citep{riffel08}. In this latter case, the X-ray emission of the AGN may have worked as a flashlight which ``illuminated'' the inflowing molecular gas. We obtained a mass inflow rate of $4\,\times\,10^{-5}\,$M$_\odot\,$yr$^{-1}$. If we use the same ratio of hot to total molecular gas as above, the inflow rate becomes of the order of a few M$_\odot\,$yr$^{-1}$. Another recent case of inflow seen in molecular gas was reported by \citet{sanchez09}, using near-IR IFU observations  of NGC\,1068 obtained with SINFONI.

The feedback from the AGN -- in the form of the mass outflow rate along the NLR and its kinetic power --  can be calculated using the ionised gas luminosity and geometry of the outflow through the expression:

\begin{equation}
\dot{M}_{\rm HII}=m_p\,N_{\rm HII}\,v\,f\,A
\end{equation}

\noindent where $m_p$ is the mass of the proton, $N_{\rm HII}$ is the ionised gas density, $v$ is the outflow velocity, $f$ is the gas filling factor and $A$ is the cross section area of the outflow. The filling factor  was estimated from:

\begin{equation}
 L_{Br\,\gamma}=j_{Br\,\gamma} V f
\end{equation}

\noindent where $L_{Br\,\gamma}$ is the Br\,$\gamma$ luminosity emitted by a volume $V$ and $j_{Br\,\gamma}=6.94\,\times10^{-23}$\,erg\,cm$^{-3}$\,s$^{-1}$ for a temperature T=10000K and gas density $N_{\rm HII}=N_e=500\,$cm$^{-3}$ \citep{osterbrok89}. \footnote{We have tried a lower density of 100\,cm$^{-3}$ as in \citetalias{sb09} but got as a result f$>$1, which is unphysical, supporting a higher average gas density.}

The volume $V$ was calculated assuming the geometry proposed by \citet{das05} and used in the modelling described in the previous section, namely a hollow bi-cone with a height of 130\,pc for each cone (corresponding to 2\arcsec), and inner and outer opening angles of  $15\deg$ and $33\deg$, respectively. We obtained $V=2.88\,\times\,10^{61}$cm$^{-3}$. The Br\,$\gamma$ luminosity was obtained from Paper\,I. 
We have further divided this by a factor of 2, because part of the ionised gas emission comes from galactic disc gas that is not outflowing (see previous section). We estimate from the channel maps that approximately half of the total ionised gas luminosity arises in this galactic disc gas. Substituting these estimates in the expression above we obtain a filling factor value of $f=0.11$.  Filling factor values reported in the literature are usually smaller, but they usually  consider an isotropic (spherical) outflow. Assuming an isotropic outflow instead of a bi-conical one, we obtain $f=0.025$ which is similar to the typical values in the literature.

The cross section area $A$ of the hollow cone was calculated at a distance of 65\,pc (1\arcsec) from the cone apex for the hollow cone geometry and resulted in $A=5\,\times\,10^{40}$cm$^2$. Adopting the velocity $v=600$\,km\,s$^{-1}$ for the outflow (which is the one assumed for the constant velocity outflow model), $N_{\rm HII}=500$\,cm$^{-3}$, $f=0.11$ we finally obtain a mass outflow rate of $\dot{M}_{\rm HII}\approx\,1.2\,$M$_\odot\,$yr$^{-1}$ in each cone. This value is of the order of the mass outflow rates reported for the NLR of active galaxies in neutral and ionised gas \citep{veilleux05,holt06}, and $\sim$6 times larger than the mass outflow rate estimated by \citet{crenkra07} from  UV absorbers in the spectra.

The mass outflow rate can be compared with the accretion rate to feed the supermassive black hole. The accretion rate can be estimated from:

\begin{equation}
\dot{m}=\frac{L_{bol}}{c^2 \eta}
\end{equation}

\noindent where $c$ is the speed of light, and the bolometric luminosity is $L_{bol}=7.3\times10^{43}\,$erg\,s$^{-1}$ \citep{kaspi05}, which is 0.012 times the Eddington luminosity for a black hole mass of $4.5\,\times\,10^7$\,M$_\odot$ \citep{onken07,bentz06}. For this Eddington ratio we adopt an efficiency factor for the conversion of rest mass energy into radiation of  $\eta=0.1$, a typical value for an optically thin geometrically thick accretion disc \citep{frank02}. The result is $\dot{m}=1.3\times10^{-2}\,M_\odot\,$yr$^{-1}$, which is two orders of magnitude smaller than the mass outflow rate, indicating that the origin of the outflowing gas is not the AGN but could be surrouding gas from the galaxy ISM accelerated by a nuclear outflow, as discussed in the previous section. 

Our data also allows us to estimate the kinetic power of the outflow, including both the radial and turbulent components of the gas motion. Assuming that the velocity dispersion of the H{\sc ii} emitting gas is due to turbulence, the kinetic power can be estimated as:

\begin{equation}
\dot{E}\approx\frac{\dot{M}_{H_{\rm HII}}}{2}(v^2+\sigma^2).
\end{equation}

\noindent  From Fig.~\ref{f-doublesig} we obtain $\sigma\approx150$\kms, and adopting as above, $v\,=\,600$\kms and using $\dot{M}_{\rm HII}\,\approx\,2.4\,{\rm \,M_\odot\,yr^{-1}}$ (considering the outflows in both cones), we obtain $\dot{E}\approx\,2.4\,\times10^{41}$\,erg\,s$^{-1}$, which is $\approx\,0.03\times$L$_{bol}$.  Thus, from the power that can be extracted from the accretion rate of  $\dot{m}=1.3\times10^{-2}\,M_\odot\,$yr$^{-1}$, $\approx$\,0.3\% is transformed in kinetic power in the NLR outflows. We note that this fraction is of the order of the AGN feedback derived in simulations for co-evolution of black holes and galaxies (0.5\%), in order to match the M-$\sigma$ relation \citep{matteo05}. This kinetic power is somewhat larger than those we have recently derived for a sample of 6 nearby Seyfert galaxies showing NLR outflows \citep{barbosa09}, which are in the range $10^{-5} \le \dot{E}/L_{bol} \le 10^{-4}$. 

In order to gauge the impact that the above kinetic power may have in the galaxy, we estimate the total kinetic energy released if the outflow as now observed lasts for at least $10^7$\,yrs (as an estimated duration for the activity phase of the galaxy) to be $7.6\,\times\,10^{55}$\,erg.  This energy can then be equated to the binding energy of the gas mass that could be pushed away from the central region of the galaxy (following the reasoning in \citet{holt06}) $E_{bind}\,\approx G M_{gas}M_{total}/R_{gas}$. For $E_{bind}=\,7.6\,\times\,10^{55}$\,erg, $M_{total}=\times 10^9$\,M$_\odot$ (estimated from the galaxy absolute magnitude) and $R_{gas}\,\approx\,500$\,pc (just beyond the observed outflow), we obtain $M_{gas}\approx\,10^8$\,M$_\odot$, which is $\approx$ 5 times the mass which outflows through the NLR in 10$^7$\,yrs. It is interesting to note that this mass is of the order of the estimated molecular gas mass present in the nuclear region of NGC\,4151, as discussed above, being also the typical value observed in the nuclear region of active galaxies in general \citep{gb05,boone07}. Thus, it appears that in the case of NGC\,4151 the molecular gas reservoir that is needed to trigger the nuclear activity can be blown away by the NLR outflow by the end of the activity cycle.




\section{Conclusions}
\label{conclusion}

We have measured the kinematics of the ionised and molecular gas surrounding the nucleus of  NGC\,4151 using integral-field spectroscopy obtained with the NIFS instrument at the Gemini North telescope. The main results of this paper are:

\begin{itemize}

\item The ionised-gas kinematics are dominated by three velocity components: (1) extended emission at systemic velocity observed in a circular region around the nucleus;  (2) a component outflowing along the bi-cone (PA$=60\deg$); (3) another component due to the interaction of the radio jet with the galactic disk.

\item Regarding the extended emission at systemic velocity: The origin of this component is gas from the galaxy plane ionised by the AGN up to at least 1 arcsec (65\,pc) from the nucleus, revealing that there is escape of ionizing radiation even along the equatorial plane of the bi-cone.

\item Regarding the outflowing component: Channel maps extracted along the emission-line profiles show high-velocity gas all the way from the nucleus to the border of the emitting region. This seems to contradict  previous studies showing acceleration along the NLR. 
Our data suggest that the NLR clouds are accelerated very close to the nucleus (within $\approx$ 10\,pc), after which the flow moves at essentially constant velocity. 

\item We do not detect NIR counterparts to the highest-velocity \foiii\ emission seen by previous authors. It may originate in very low-density gas, which could be ablated material flowing off NLR clouds. This emission would not be characteristic of the bulk of the NLR mass.

\item The origin of the bulk of NLR mass seems to be  entrained gas from the galaxy plane. This interpretation is supported by the large mass-outflow rate in ionised gas of $\sim$\,1\,M$_\odot$\,yr$^{-1}$ obtained for the outflowing component in each cone, which is $\sim$100 times the nuclear mass accretion rate.

\item The estimated kinetic power of the outflow is $\approx2.4\times10^{41}$\,erg\,s$^{-1}$, which is $\sim$\,0.3\% of L$_{bol}$, and is $\approx$\,10 times larger than values obtained previously by our group for other nearby Seyfert galaxies.

\item Regarding the third kinematic component: Our data clearly show the interaction of the radio jet with ambient gas at systemic velocity. This has not been seen previously and indicates that the jet is launched close to the plane of the galaxy, pushing the gas encountered on its way.

\item  Molecular hydrogen emission arises in extended regions along the axis of the galaxy's stellar bar, avoiding the region of the bi-cone. The \hmol\ velocities are close to systemic with a small rotation component, supporting an origin in the galaxy plane. The emitting molecular gas is probably just the ``hot skin'' of a larger reservoir that may have been built up by the previously observed HI inflow along the bar of the galaxy, and is probably the source of the AGN feeding.

\section*{Acknowledgments}
We acknowledge the suggestions by an anonymous referee which helped to improve the paper,
as well as valuable discussions with H. R. Schmitt, S. Kraemer, M. Crenshaw and V. Das.
RDSL thanks the hospitality of the Department of Astronomy and Center for Cosmology and Astroparticle Physics of The Ohio State University, where part of this work was done.
Based on observations obtained at the Gemini Observatory, which is
operated by the Association of Universities for Research in Astronomy,
Inc., under a cooperative agreement with the NSF on behalf of the
Gemini partnership: the National Science Foundation (United States),
the Science and Technology Facilities Council (United Kingdom), the
National Research Council (Canada), CONICYT (Chile), the Australian
Research Council (Australia),  Minist\'erio da Ci\^encia e Tecnologia (Brazil) and SECYT
(Argentina). This work has been partially supported by the Brazilian
institution CNPq.

\end{itemize}


\label{lastpage}


\begin{thebibliography}{}

\bibitem[\protect\citeauthoryear{Antonucci \& Miller}{1985}]{antmi85} Antonucci, R. R. \& Miller, J. S., 1985, ApJ, 297, 621

\bibitem[\protect\citeauthoryear{Barbosa et al.}{2009}]{barbosa09} Barbosa F.~K.~B., Storchi-Bergmann T., Fernandes R.~Cid, Winge C., Schmitt H., 2009, MNRAS, 396, 2

\bibitem[\protect\citeauthoryear{Bentz et al.}{2006}]{bentz06} Bentz, M., 2006, ApJ, 651, 775

\bibitem[\protect\citeauthoryear{Boone et al.}{2007}]{boone07} Boone F., Baker A.~J., Schinnerer E. et al., 2007, A\&A, 471, 113

\bibitem[\protect\citeauthoryear{Cecil et al.}{2002}]{cecil02} Cecil G., Dopita M.~A., Groves B., Wilson A.~S., Ferruit P., P\'econtal E., Binette L., 2002, ApJ, 568, 627

\bibitem[\protect\citeauthoryear{Crenshaw \& Kraemer}{2000}]{crenshaw00b} Crenshaw D.~M., Kraemer S.~B., 2000, ApJ, 532, L101

\bibitem[\protect\citeauthoryear{Crenshaw \& Kraemer}{2007}]{crenkra07} Crenshaw D.~M. \&  Kraemer S.~B., 2007, ApJ, 659, 250

\bibitem[\protect\citeauthoryear{Crenshaw et al.}{2000}]{crenshaw00a} Crenshaw D.~M., Kraemer S.~B., Hutchings J.~B et al., 2000, AJ, 120, 1731

\bibitem[\protect\citeauthoryear{Dale et al.}{2005}]{dale05} Dale D.~A., Sheth K., Helou G., Regan M.~W., H\"uttemeister S., ApJ, 129, 2197

\bibitem[\protect\citeauthoryear{Das et al.}{2005}]{das05} Das V., Crenshaw D.~M., Hutchings J.~B. et al., 2005, AJ, 130, 945

\bibitem[\protect\citeauthoryear{Das et al.}{2006}]{das06} Das V., Crenshaw D.~M., Kraemer S.~B. et al., 2006, AJ, 132, 620

\bibitem[\protect\citeauthoryear{Das, Crenshaw, \& Kraemer}{2007}]{das07} Das V., Crenshaw D.~M., Kraemer S.~B., 2007, ApJ, 656, 699

\bibitem[\protect\citeauthoryear{Di Matteo et al.}{2005}]{matteo05} Di Matteo, T., Springel, V \& Hernquist, L., 2005, Nature, 433, 604


\bibitem[\protect\citeauthoryear{Evans et al.}{1993}]{evans93} Evans I.~N., Tsvetanov Z., Kriss G.~A., Ford H.~C., Caganoff S., Koratkar A.~P., ApJ, 1993, 417, 82

\bibitem[\protect\citeauthoryear{Elitzur \& Shlosman}{2006}]{elitzur06} Elitzur, M. \& Shlosman, I., 2006, ApJ, 648, L101

\bibitem[\protect\citeauthoryear{Everett \& Murray}{2007}]{everett07} Everett J.~E., Murray N., 2007, ApJ, 656, 93

\bibitem[\protect\citeauthoryear{Fathi et al.}{2006}]{fathi06} Fathi K., Storchi-Bergmann T., Riffel R.~A., Winge C., Axon D.~J.,  Robinson A., Capetti A., Marconi A., 2006, ApJ, 641, L25

\bibitem[\protect\citeauthoryear{Ferland et al.}{1998}]{ferland98} Ferland G.~J., Korista K.~T., Verner D.~A., Ferguson J.~W., Kingdon J.~B., Verner E.~M., 1998, PASP, 110, 761

\bibitem[\protect\citeauthoryear{Frank et al.}{2002}]{frank02} Frank J., King A.~R., Raine D.~J. 2002, Accretion Power in Astrophysics (3rd ed., Cambridge: Cambridge Univ. Press)

\bibitem[\protect\citeauthoryear{Garc\'ia-Burillo et al.}{2005}]{gb05} Garc\'ia-Burillo S., Combes F., Schinnerer E., Boone F., Hunt L.~K., 2005, A\&A, 441, 1011

\bibitem[\protect\citeauthoryear{Holt et al.}{2006}]{holt06} Holt J., Tadhunter C., Morganti R., Bellamy M., Gonz\'alez Delgado R.~M., Tzioumis A., Inskip K.~J., 2006, MNRAS, 370, 1633

\bibitem[\protect\citeauthoryear{Hutchings et al.}{1998}]{hutchings98} Hutchings J.~B., Crenshaw D.~M., Kaiser M.~E. et al., 1998 ApJ, 492, L115

\bibitem[\protect\citeauthoryear{Hutchings et al.}{1999}]{hutchings99} Hutchings J.~B., Crenshaw D.~M., Danks A.~C. et al., 1999, AJ, 118, 2101

\bibitem[\protect\citeauthoryear{Kaiser et al.}{2000}]{kaiser00} Kaiser M.~E., Bradley L.~D.~II, Hutchings J.~B. et al., 2000, ApJ, 528, 260

\bibitem[\protect\citeauthoryear{Kaspi et al.}{2005}]{kaspi05} Kaspi, S., Maoz, D., Netzer, H., Peterson, B. M. Vestergaard, M. \& Jannuzi, B. T., 2005, ApJ, 629, 61

\bibitem[\protect\citeauthoryear{Kraemer et al.}{2008}]{kra08} Kraemer, S. B., Schmitt, H. R. \&  Crenshaw, D. M. 2008, ApJ, 679, 1128

\bibitem[\protect\citeauthoryear{Krips et al.}{2007}]{krips07} Krips M., Neri R., Garc\'ia-Burillo S. et al., 2007, A\&A, 468, L63

\bibitem[\protect\citeauthoryear{Lada \& Fich}{1996}]{lada96} Lada C.~J., Fich M., 1996, ApJ, 459, 638

\bibitem[\protect\citeauthoryear{Matzner \& McKee}{1999}]{matzner99} Matzner C.~D., McKee C.~F., 1999, ApJ, 526, L109

\bibitem[\protect\citeauthoryear{McGregor et al.}{2003}]{mcgregor03} McGregor P. J. et al., 2003, in Iye M., Moorwood, A.F.M, eds, Proc. SPIE Vol. 4841, Instrument Design and Performance for Optical/Infrared Ground-based Telescopes. SPIE, p. 1581

\bibitem[\protect\citeauthoryear{McGregor et al.}{2010}]{McGregor10} McGregor P. J. et al., 2010, work in progress

\bibitem[\protect\citeauthoryear{S\'anchez et al.}{2009}]{sanchez09} S\'anchez F.~M, Davies R.~I., Genzel R., Tacconi L.~J., Eisenhauer F., Hicks E.~K.~S., Friedrich S., Sternberg A., 2009, ApJ, 691, 749

\bibitem[\protect\citeauthoryear{Mundell \& Shone}{1999}]{mundell99} Mundell C.~G., Shone D.~L., 1999, MNRAS, 304, 475

\bibitem[\protect\citeauthoryear{Mundell et al.}{1995}]{mundell95} Mundell C.~G., Pedlar A., Baum S.~A., O'Dea C.~P., Gallimore J.~F., Brinks E., 1995, MNRAS, 272, 355

\bibitem[\protect\citeauthoryear{Mundell et al.}{2003}]{mundell03} Mundell C.~G., Wrobel J.~M., Pedlar A., Gallimore J.~F., 2003, ApJ, 583, 192

\bibitem[\protect\citeauthoryear{Onken et al.}{2007}]{onken07} Onken, C. A., 2007, ApJ, 670, 1050

\bibitem[\protect\citeauthoryear{Osterbrok}{1989}]{osterbrok89} Osterbrock, D., 1989, Astrophyscis of Gaseous Nebulae and Active Galactic Nuclei, University Science Books

\bibitem[\protect\citeauthoryear{Pedlar et al.}{1992}]{pedlar92} Pedlar A., Howley P., Axon D.~J, Unger S.~W., 1992, MNRAS, 259, 369

\bibitem[\protect\citeauthoryear{Pedlar et al.}{1993}]{pedlar93} Pedlar A., Kukula M.~J., Longley D.~P.~T., Muxlow T.~W.~B., Axon D.~J., Baum S., O'Dea C., Unger S.~W, 1993, MNRAS, 263, 471P


\bibitem[\protect\citeauthoryear{Riffel et al.}{2006}]{riffel06} Riffel R.~A., Storchi-Bergmann T., Winge C., Barbosa F.~K.~B., 2006, MNRAS, 373, 2

\bibitem[\protect\citeauthoryear{Riffel et al.}{2008}]{riffel08} Riffel R.~A., Storchi-Bergmann T., Winge C., McGregor P.~J., Beck T., Schmitt H., 2008, MNRAS, 385, 1129

\bibitem[\protect\citeauthoryear{Riffel et al.}{2009a}]{riffel09} Riffel R.~A., Storchi-Bergmann T., Dors O.~L.~Jr, Winge C., 2009, MNRAS, 393, 783

\bibitem[\protect\citeauthoryear{Riffel et al.}{2009b}]{riffel09b} Riffel R.~A., Storchi-Bergmann T., McGregor, P. J., 2009, ApJ, 698, 1767

\bibitem[\protect\citeauthoryear{Rodr\'iguez-Ardila et al.}{2005}]{ardila05} Rodr\'iguez-Ardila A., Riffel R., Pastoriza M. G., 2005, MNRAS, 364, 1041

\bibitem[\protect\citeauthoryear{Schulz}{1990}]{schulz90} Schulz H., 1990, AJ, 99, 1442

\bibitem[\protect\citeauthoryear{Simkin}{1975}]{simkin75} Simkin S.~M., 1975, ApJ, 200, 567S

\bibitem[\protect\citeauthoryear{Storchi-Bergmann, Kinney, \& Challis}{1995}]{sb95} Storchi-Bergmann T., Kinney A.~L., Challis P., 1995, ApJS, 98,
103

\bibitem[\protect\citeauthoryear{Storchi-Bergmann et al.}{1999}]{sb99} Storchi-Bergmann T., Winge C., Ward M.~J., Wilson A.~S., 1999, MNRAS, 304, 35

\bibitem[\protect\citeauthoryear{Storchi-Bergmann et al.}{2007}]{sb07} Storchi-Bergmann T., Dors O.~L.~Jr., Riffel R.~A., Fathi K., Axon D.~J., Robinson A., Marconi A., \"Ostlin G., 2007, ApJ, 670, 959

\bibitem[\protect\citeauthoryear{Storchi-Bergmann et al.}{2009}]{sb09} Storchi-Bergmann T., McGregor P.~J., Riffel R.~A., Sim\~oes Lopes R.~D., Beck T., Dopita M., 2009, MNRAS, 394, 1148 (Paper I)

\bibitem[\protect\citeauthoryear{Veilleux et al.}{2005}]{veilleux05} Veilleux S., Cecil G., Bland-Hawthorn J., 2005, ARA\&A, 43, 769

\bibitem[\protect\citeauthoryear{Wang et al.}{2009}]{wang09} Wang, J.,  Fabbiano, G., Karovska, M., Elvis, M., Risaliti, G., Zezas, A.,  Mundell, C. G. 2009, ApJ, 704, 1195

\bibitem[\protect\citeauthoryear{Winge et al.}{1997}]{winge97} Winge C., Axon D.~J., Macchetto F.~D., Capetti A., 1997, ApJ, 487, L121

\end{thebibliography}
\end{document}